\documentclass[12pt,draftclsnofoot, onecolumn]{IEEEtran}

\ifCLASSINFOpdf
\else
\fi

\usepackage{amsmath,graphicx}
\usepackage{subfigure}
\usepackage{lineno}
\usepackage{cite}
\usepackage{multirow}
\usepackage{stfloats}
\usepackage{amsfonts}
\usepackage{amssymb}
\usepackage{booktabs} 
\usepackage{makecell}
\usepackage{diagbox}
\usepackage{slashbox}
\usepackage{color}
\usepackage{lipsum}
\usepackage{amssymb}
\usepackage{bbding}
\usepackage[ruled,linesnumbered,boxed]{algorithm2e}
\begin{document}
	
	\title{Algorithm Unrolling-Based \\ Distributed Optimization for RIS-Assisted \\Cell-Free Networks}
	
	\author{
		\IEEEauthorblockN{
			Wangyang Xu, Jiancheng An, \IEEEmembership{Member, IEEE}, Hongbin Li, \IEEEmembership{Fellow, IEEE}, \\Lu Gan, and Chau Yuen, \IEEEmembership{Fellow, IEEE}
			\vspace{-3em}}
		\thanks{W. Xu and L. Gan are with the School of Information and Communication Engineering, University of Electronic Science and Technology of China, Chengdu, Sichuan, 611731, China (e-mail: wangyangxu@std.uestc.edu.cn; ganlu@uestc.edu.cn).}
		\thanks{H. Li is with the Department of Electrical and Computer Engineering, Stevens Institute of Technology, Hoboken, NJ 07030, USA (e-mail: hli@stevens.edu).}
		\thanks{J. An and C. Yuen are with Engineering Product Development (EPD) Pillar, Singapore University of Technology and Design, Singapore 487372, Singapore (e-mail: jiancheng\_an@sutd.edu.sg; yuenchau@sutd.edu.sg).}
	}
\maketitle

\begin{abstract}
The user-centric cell-free network has emerged as an appealing technology to improve the next-generation wireless network's capacity thanks to its ability to eliminate inter-cell interference effectively. However, the cell-free network inevitably brings in higher hardware cost and backhaul overhead as a larger number of base stations (BSs) are deployed. Additionally, severe channel fading in high-frequency bands constitutes another crucial issue that limits the practical application of the cell-free network. In order to address the above challenges, we amalgamate the cell-free system with another emerging technology, namely reconfigurable intelligent surface (RIS), which can provide high spectrum and energy efficiency with low hardware cost by reshaping the wireless propagation environment intelligently. To this end, we formulate a weighted sum-rate (WSR) maximization problem for RIS-assisted cell-free systems by jointly optimizing the BS precoding matrix and the RIS reflection coefficient vector. Subsequently, we transform the complicated WSR problem to a tractable optimization problem and propose a distributed cooperative alternating direction method of multipliers (ADMM) to fully utilize parallel computing resources. Inspired by the model-based \textit{algorithm unrolling} concept, we unroll our solver to a learning-based deep distributed ADMM (D$^2$-ADMM) network framework. To improve the efficiency of the D$^2$-ADMM in distributed BSs, we develop a monodirectional information exchange strategy with a small signaling overhead. In addition to benefiting from domain knowledge, D$^2$-ADMM adaptively learns hyper-parameters and non-convex solvers of the intractable RIS design problem through data-driven end-to-end training. Finally, numerical results demonstrate that the proposed D$^2$-ADMM achieve around $210\%$ improvement in capacity compared with the distributed noncooperative algorithm and almost $96\%$ compared with the centralized algorithm.
\end{abstract}


%
\IEEEpeerreviewmaketitle

\section{Introduction}
%
%
%
%
Next-generation wireless communication systems are expected to meet an even greater demand for higher capacity, denser connectivity, and broader coverage with the advent of the internet of everything \cite{B5G, B5G2, 6G, hongbinli2}. The conventional communication network relies on cellular topology where effective communication paradigms, such as  small-cell network and cellular massive multiple-input multiple-output (MIMO) are developed based on cell-centric principles \cite{smallcell, cellularMIMO}. Specifically, a single base station (BS) serves all users in the same cell while appropriate resource reuse policies are adopted among different cells. As a result, users at the cell edge are more likely to be disturbed by the uplink/downlink signals from other adjacent cells, resulting in the common issue of \emph{inter-cell interference} \cite{cellfree1}.

It has been demonstrated that small-cell network can achieve better energy efficiency than cellular massive MIMO in some typical scenarios by properly reducing the cell size \cite{comsmacel1, comsmacel2}. However, as the cell density increases, the inter-cell interference will increase accordingly and become the main bottleneck limiting the capacity of the cellular network \cite{comsmacel4}. Although cellular massive MIMO is not affected by the inter-cell interference, the shadow fading due to blocking will become a performance-limiting factor  if a large number of antennas are centralizedly configured on a single BS. Therefore, cellular massive MIMO's coverage and network capacity may be significantly deteriorated in some harsh environments \cite{comsmacel3}.

In sharp contrast to the aforementioned cell-centric networks, a user-centric network paradigm known as the \emph{cell-free massive MIMO network} has recently received significant attention as a potential and cutting-edge substitute \cite{cellfree1, cellfree2, cellfree3}. In a cell-free massive MIMO network, a large number of antennas is spread on numerous BSs in a distributed form \cite{cellfree4}. These BSs provide service to a relatively small number of users within the same time-frequency domain. Since cell-free massive MIMO removes the underlying cell edge, it does not cause inter-cell interference as existing cellular networks. Although cell-free massive MIMO has many appealing advantages, its application in higher frequency bands of future communication systems still has to overcome issues related to severe transmission attenuation and coverage blind spots \cite{highfreband1, highfreband2}. In addition, deploying a large number of BSs also brings prohibitive hardware cost and energy consumption. Fortunately, a promising technology, reconfigurable intelligent surface (RIS), has recently been introduced in various communication scenarios to significantly improve the system throughput and spectrum/energy efficiency \cite{1WQQtoward2020, huangTWC, xuwangyang, an3}. Specifically, an RIS is a metal panel equipped with many low-cost passive  elements. The phase shifts of these passive elements can be adjusted to achieve intelligent manipulation of the wireless environment and enhance communication quality-of-service (QoS) \cite{xuwangyang1}. In view of the superiority of the RIS and cell-free systems, it is of interest to develop an RIS-assisted cell-free approach for future wireless communications.

\subsection{Prior Works}
Downlink precoding is crucial to unleash the full potential of cell-free networks. Currently, most existing precoding schemes for cell-free systems can be generally classified into non-cooperative \cite{cellfree1, cellfreenoncopZF, cellfreenoncopMMSE} and cooperative \cite{cellfreecopZF1, cellfreecopZF2, cellfree3, hongbinli1}. Non-cooperative precoding assumes that each BS can only utilize local channel state information (CSI) acquired through uplink channel estimation, without performing any CSI exchange among BSs. Along this direction, some rather simple strategies such as maximum ratio transmission (MRT) \cite{cellfree1}, local zero-forcing (ZF) \cite{cellfreenoncopZF}, and local minimum mean square error (MMSE) \cite{cellfreenoncopMMSE} designs have been employed for precoding. Cooperative precoding including both centralized and distributed cooperative schemes that perform joint precoding across all BSs achieves better system performance compared with its non-cooperative counterpart. Specifically, in the centralized scheme, BSs upload their local CSI to the centralized processing unit (CPU) through a specific backhaul link, based on which the precoding matrices of all BSs are jointly designed and then distributed. Most existing works on centralized precoding concentrate on developing precoding algorithms for the CPU, such as the centralized ZF precoding \cite{cellfreecopZF1, cellfreecopZF2} and the centralized MMSE precoding \cite{cellfree3}. Distributed cooperative precoding distributes the computational load to multiple BSs, thus reducing the computational burden of the CPU \cite{cellfreecd1, cellfreecd2}. The precoding of each BS is carried out locally and updated based on the cross-term information exchange among different BSs to approach the optimal performance of the  centralized design.

Meanwhile, RIS reflection coefficient design has received much attention recently, e.g., \cite{wujointcontinuous, an1, huang2020reconfigurable, huang2020holographic}, by taking into account of various practical constraints and application backgrounds. However, the research on RIS-assisted cell-free network is still in its infancy stage \cite{RIScellfree1, RIScellfree2, RIScellfree3, RIScellfree4, RIScellfree5}. Specifically, the authors of \cite{RIScellfree2} used the conjugate beamforming method with the local CSI to design precoding vectors along with randomly adjusted RIS reflection coefficient vector to illustrate the performance gain of the cell-free system. By assuming that BSs send their local CSI to the CPU, the authors of \cite{RIScellfree3, RIScellfree4} adopted alternating optimization algorithms to jointly design the BS precoding and the RIS reflection coefficient vector. Note that the works mentioned above are based on either a non-cooperative scheme, which requires no CSI exchange but yields inferior performance, or a centralized scheme, which trades system complexity for better performance. Although recently \cite{RIScellfree5} proposed a distributed cooperative optimization method for RIS-assisted cell-free system, it has to perform a set of iterations at different BSs without fully taking advantage of the distributed parallel computing capabilities of cell-free systems.

Also, most existing research efforts on RIS-assisted cell-free systems are focused on developing iterative optimization algorithms, which are based on some sophisticated models derived from the underlying  physical processes or through handcrafting \cite{antgcn, anwc}. On the contrary, deep learning (DL) methods attempt to automatically infer model information and network parameters directly from training data \cite{xuwangyang, xu3}. Therefore, DL is very promising for scenarios where the environment is complex and the system model is challenging to be constructed explicitly \cite{deepphy}. In addition, the number of layers of most neural networks is much fewer than the number of iterations incurred by typical iterative algorithms, which allows DL methods to attain a faster inference speed. Nevertheless, neural networks are often trained as a ``black-box'' with poor interpretability and lack essential domain knowledge that is beneficial for generalization. Therefore, combining conventional iterative algorithms and raw data-driven DL has become a new surge of research. Recently, an appealing concept called \textit{algorithm unrolling} has been proposed, which unrolls iteration-based algorithms into learning-based neural network structures \cite{Alunroll1, Alunroll2, Alunroll3, Alunroll4, Alunroll5, Alunroll6, Alunroll7, Alunroll8, Alunroll9, Alunroll10}. Such a unfolding process can not only integrate domain knowledge but also learn complex mapping functions and hyper-parameters from input data. Specifically, each step in the traditional iterative algorithm is unrolled into a layer or a block of the neural network. Different network layers or blocks are cascaded to form a holistic neural network framework for solving the original problem more efficiently. The algorithm unrolling methods have shown advantages in many application domains, such as computational imaging \cite{Alunroll7, Alunroll8}, speech processing \cite{Alunroll9}, and remote sensing \cite{Alunroll10}.
\begin{table*}[t]
	\centering
	\caption{{Contrast of the Proposed D$^2$-ADMM to Existing Works for RIS-Assisted Cell-Free Systems}}
	\begin{tabular}{|l|l|c|c|c|c|c|}
		\hline
		\multicolumn{2}{|l|}{Features} & Proposed & \cite{RIScellfree2} & \cite{RIScellfree3} & \cite{RIScellfree4} & \cite{RIScellfree5}\\
		\hline
		
		\multicolumn{2}{|l|}{Optimization objective}  &  WSR & ASR & EE & WSR & WSR\\
		\hline
		
		\multicolumn{2}{|l|}{BS precoding design}  &  DL & Local MRT & IA & PDS & ADMM\\
		\hline
		
		\multicolumn{2}{|l|}{RIS passive beamforming design} &  DL & Random & IA & PDS & MM\\
		\hline
		
		\multicolumn{2}{|l|}{Centralized design} &$\times$	&$\times$	&\checkmark	&\checkmark	&$\times$\\ 
		\hline
		
		\multirow{2}{*}{Distributed design} & Noncooperative & $\times$ & \checkmark & $\times$ & $\times$ & $\times$ \\
		\cline{2-7}
		
		\multirow{2}{*}{} & Cooperative & \checkmark & $\times$ & $\times$ & $\times$ & \checkmark \\
		\hline
		
		\multicolumn{2}{|l|}{Convergence speed} &Fast & N/A & Moderate & Moderate & Slow\\ 
		\hline
	\end{tabular}\\
	\vspace{0.25cm}
	WSR: weighted sum-rate; ASR: average sum-rate; EE: energy efficiency; IA: inner approximation; \\
	PDS: primal-dual subgradient; MM: majorization-minimization;\\
	\hrulefill
	\label{contribution}
\end{table*}

\subsection{Contributions}
Targeting RIS-assisted cell-free systems, we design a fully distributed joint BS precoding and RIS reflection coefficient optimization scheme based on the alternating direction method of multipliers (ADMM). Furthermore, we unroll the proposed solver to a learning-based neural network to attain better convergence and system performance. More specifically, the main contributions of this paper in contrast with existing works are shown in Table \ref{contribution} and further summarized as follows:
\begin{itemize}
\item We propose a distributed RIS-assisted cell-free system, where multiple energy-efficient RISs are deployed to assist in the downlink communications from a set of distributed BSs to multiple users (UEs). A distributed cooperative BS precoding and RIS reflection coefficient design scheme is developed to make full use of distributed computing resources.  
\item Furthermore, we propose a distributed design based on ADMM that iteratively updates the corresponding auxiliary variables, BS precoding, RIS reflection coefficient vectors, and multipliers involved. The proposed design considers the consensus problem when separately designing the RIS reflection coefficients at each BS in parallel. 
\item We unroll the proposed distributed ADMM design into a learning-based deep distributed ADMM (D$^2$-ADMM) neural network structure, which consists of a cascade of multiple neural blocks. Each neural block is designed by unfolding a single iteration of the proposed distributed ADMM design.  Moreover, an effective monodirectional information exchange strategy with a small information exchange overhead is proposed for implementing our algorithm. In addition to obtaining deterministic variable updating strategies from domain knowledge, D$^2$-ADMM adaptively learns hyper-parameters and non-convex solvers of the RIS design problem through data-driven end-to-end training. Furthermore, D$^2$-ADMM requires only a few neural blocks to reach convergence thanks to the strong inferential capability of DL.
\item Finally, we elaborate on the training and implementation of the proposed algorithm. Numerical results demonstrate that the proposed algorithm has faster convergence, less computational complexity, and better performance compared with various traditional algorithms. 
\end{itemize}
\subsection{Organization and Notations}
The rest of this paper is organized as follows. Section II introduces the system model and formulates the joint precoding and RIS reflection design problem in RIS-assisted cell-free systems. In Sections III, we propose a distributed ADMM-based design by maximizing the weighted sum-rate. Section IV presents a D$^2$-ADMM neural network structure and a monodirectional information exchange strategy to design the BS precoding and the RIS reflection coefficients. Numerical results are provided in Section V. Finally, we conclude the paper in Section VI.

\emph{Notations:} In this paper, scalars are denoted by italic letters. Vectors and matrices are denoted by bold-face lower-case and upper-case letters, respectively. The superscripts ${(  \cdot  )^T}$ and ${(  \cdot  )^H}$ represent the operations of transpose and Hermitian transpose. $\left|  \cdot  \right|$ denotes the absolute value of a real number. $\left\|  \cdot  \right\|$ denotes the 2-norm of a vector or a matrix. $\operatorname{Re} \left\{  x  \right\}$ and $\operatorname{Im}  \left\{  x  \right\}$ denote the real and imaginary parts of the complex number $x$, respectively. ${\rm{diag}}\left(  \cdot  \right)$ denotes the diagonal operation. The distribution of a circularly symmetric complex Gaussian (CSCG) with mean $v$ and variance $\sigma$ is denoted as $\sim {\cal C}{\cal N}(v,\sigma)$. ${\rm log}_2(\cdot)$ represents the logarithmic function. ${{\mathbb{C}}}$ denotes the set of complex values. ${{\mathbb{S}}}$ denotes the set of symmetric positive definite matrices. 

\section{System Model and Problem Formulation}
This section starts by introducing the system model of the RIS-assisted cell-free system. In order to design the BS precoding and RIS reflection coefficient vector, a practical weighted sum-rate (WSR) maximization problem is formulated.
\subsection{System Model}
\begin{figure}[tbp]
\centering {
	\begin{tabular}{ccc}
		\includegraphics[width=0.6\textwidth]{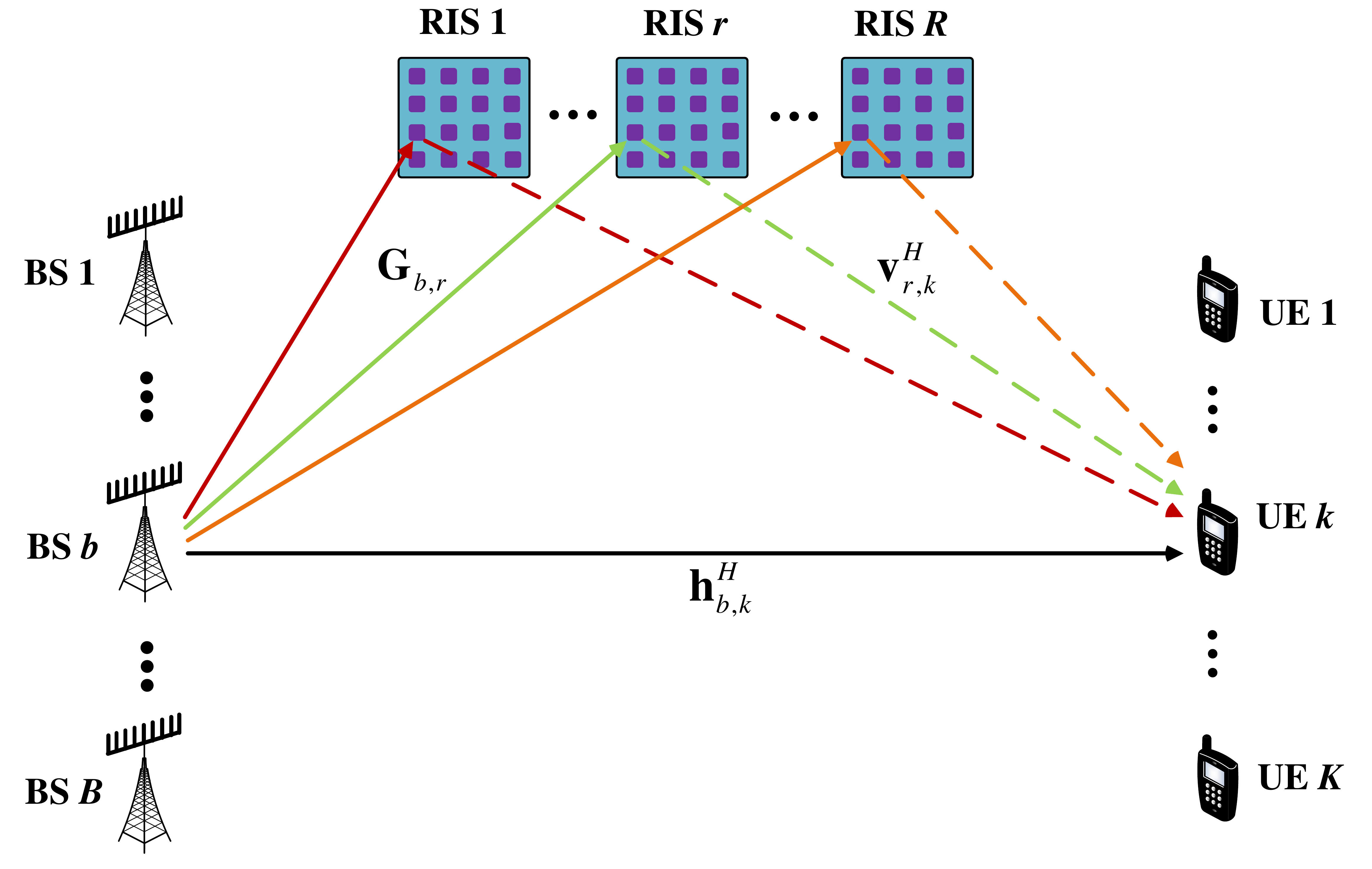}\\
	\end{tabular}
}
\caption{The downlink RIS-assisted cell-free system.}
\vspace{-1\baselineskip}
\label{CFsystem}
\end{figure}

In this paper, we consider a downlink RIS-assisted cell-free system, as illustrated in Fig. \ref{CFsystem}, where multiple BSs and RISs are deployed in a distributed arrangement to serve all UEs cooperatively. The sets of BSs, RISs, and UEs are defined as $\mathcal{B} = \{ {1,2, \cdots B} \}$, $\mathcal{R} = \{ {1,2, \cdots R} \}$, and $\mathcal{K} = \{ {1,2, \cdots K} \}$, respectively. The number of antennas of each BS and UE is $N_t$ and $1$, respectively. Each RIS is equipped with a rectangular metasurface having $N$ passive reflecting elements. 

As shown in Fig. \ref{CFsystem}, each RIS builds one virtual channel consisting of the BS-RIS and RIS-UE channels between a BS and a UE to assist in the downlink communication. In this paper, the BS-RIS channel between the $b$-th BS and the $r$-th RIS is denoted by ${{\bf{G}}_{b,r}} \in {\mathbb{C}^{N \times {N_t}}}$. The RIS-UE channel between the $r$-th RIS and the $k$-th UE is denoted by ${\bf{v}}_{r,k}^H \in {\mathbb{C}^{1 \times N}}$. Moreover, the direct channel between the $b$-th BS and the $k$-th UE is denoted by ${\bf{h}}_{b,k}^H \in {\mathbb{C}^{1 \times {N_t}}}$. We consider that the proposed system operates in the mmWave band, where ${\bf{G}}_{b,r}$, ${\bf{v}}_{r,k}$, and ${\bf{h}}_{b,k}$ are described by the Saleh-Valenzuela model \cite{wangSVmodel}, which are expressed as 
\begin{align}   \label{eq1}
{{\bf{G}}_{b,r}} &= \sqrt {\frac{{{N_t}N}}{{{L_{\bf{G}}}}}} \sum\limits_{l = 1}^{{L_{\bf{G}}}} {{\beta _{b,r,l}}} {{\bf{a}}_P}\left( {{\psi _{b,r,l}},{\varsigma _{b,r,l}}} \right){{\bf{a}}^{H}_L}\left( {{\chi _{b,r,l}}} \right),\notag\\
{{\bf{v}}_{r,k}} &= \sqrt {\frac{N}{{{L_{\bf{v}}}}}} \sum\limits_{l = 1}^{{L_{\bf{v}}}} {{\beta _{r,k,l}}} {{\bf{a}}_P}\left( {{\psi _{r,k,l}},{\varsigma _{r,k,l}}} \right),\notag\\
{{\bf{h}}_{b,k}} &= \sqrt {\frac{{{N_t}}}{{{L_{\bf{h}}}}}} \sum\limits_{l = 1}^{{L_{\bf{h}}}} {{\beta _{b,k,l}}} {{\bf{a}}_L}\left( {{\psi _{b,k,l}}} \right)
\end{align}
respectively, where ${{L_{\bf{G}}}}$, ${{L_{\bf{v}}}}$, and ${{L_{\bf{h}}}}$ denote the multi-path number of ${{\bf{G}}_{b,r}}$, ${{\bf{v}}_{r,k}}$, and ${{\bf{h}}_{b,k}}$, respectively. ${{\psi _{{\bf{*}},{\bf{*}},l}}}({{\varsigma _{{\bf{*}},{\bf{*}},l}}})$, and ${{\chi _{{\bf{*}},{\bf{*}},l}}}$  denote the azimuth (elevation) angles of arrival (AoAs), and azimuth angles of departure (AoDs), where ${\bf{*}}$ represents the index of the BS, the RIS element, and the UE, respectively. ${\beta _{b,r,l}} \sim {\cal C}{\cal N}(0,PL_{b,r,l})$, ${\beta _{r,k,l}} \sim {\cal C}{\cal N}(0,PL_{r,k,l})$, and ${\beta _{b,k,l}} \sim {\cal C}{\cal N}(0,PL_{b,k,l})$ denote the corresponding complex-valued path gain, where $PL_{}$ represents the path loss.  Besides, ${{\bf{a}}_L}\left( \varsigma  \right)$ and ${\bf{a}}\left( {\psi ,\varsigma } \right)$ denote the array response vectors of uniform linear array (ULA) and  uniform planar array (UPA), which are defined as
\begin{align}   \label{loc3} 
	{{\bf{a}}_L}\left( \psi  \right) = \frac{1}{{\sqrt {{N_L}} }}{[1, \cdot  \cdot  \cdot ,{e^{j\pi {n_l}\sin \psi }}, \cdot  \cdot  \cdot ,{e^{j\pi \left( {{N_L} - 1} \right)\sin \psi }}]^T},
\end{align}
\begin{equation}   \label{eq2}
\begin{array}{l}
	{{\bf{a}}_P}\left( {\psi ,\varsigma } \right) = \frac{1}{{\sqrt {{N_x}{N_y}} }}\left[ {1, \cdot  \cdot  \cdot ,{e^{j\pi \left( {{n_x}\sin \psi \sin\varsigma  + {n_y}\cos \varsigma } \right)}}, \cdot  \cdot  \cdot ,} \right.	{\left. {{e^{j\pi \left( {\left( {{N_x} - 1} \right)\sin \psi \sin\varsigma  + \left( {{N_y} - 1} \right)\cos \varsigma } \right)}}} \right]^T},
\end{array}
\end{equation}
respectively, where $N_L$ and $n_l$ denote the total antenna number and the antenna index of ULA; $N_x$, $N_y$, $n_x$, and $n_y$ represent the horizontal antenna number, the vertical antenna number,  the horizontal antenna index, and the vertical antenna index of UPA, respectively.

In the downlink transmission, the transmitted symbol ${{\bf{x}}_b} \in {\mathbb{C}^{N_t \times 1}}$ at the $b$-th BS is defined as
\begin{align}   \label{eq3}
{{\bf{x}}_b} = \sum\limits_{k = 1}^K {{{\bf{w}}_{b,k}}{s_k}}, \; b \in \cal B,
\end{align}
where $s_k$ is the transmitted symbol for the $k$-th UE. Thus, we have ${\bf{s}}={\left[ {{s_1},{s_2}, \cdots ,{s_K}} \right]^T} \in {\mathbb{C}^{K \times 1}}$ representing the transmitted symbol vector that satisfies $\mathbb{E}[ {{\bf{s}}{{\bf{s}}^H}}] = {{\bf{I}}_K}$;  ${\bf{w}}_{b,k} \in {\mathbb{C}^{N_t \times 1}}$ denotes the precoding vector at the $b$-th BS for the $k$-th UE. 

At each UE, the received signal component corresponding to one BS includes two parts, one is that directly propagated from the BS to the UE, while the other is that superimposing mutiple signal copies reflected by $R$ RISs. Hence, the received signal component at the $k$-th UE corresponding to the $b$-th BS can be expressed as 
\begin{align}   \label{eq4}
{{y}_{b,k}} &= \left( {{\bf{h}}_{b,k}^H + \sum\limits_{r = 1}^R {{\bf{v}}_{r,k}^H{\boldsymbol{\Theta}} _r^H{{\bf{G}}_{b,r}}} } \right)\sum\limits_{k = 1}^K {{{\bf{w}}_{b,k}}{s_k}}\notag\\
&= \left( {{\bf{h}}_{b,k}^H + {{\boldsymbol{\theta}} ^H}{\bf{V}}_k^H{{\bf{G}}_b}} \right){{\bf{x}}_b}\notag\\
&= {\bf{\hat h}}_{b,k}^H{{\bf{x}}_b},
\end{align}
where ${{\boldsymbol{\Theta}} _r} = {\rm{diag}}({{{{[{e^{j{\varphi _{r,1}}}},{e^{j{\varphi _{r,2}}}}, \cdots ,{e^{j{\varphi _{r,N}}}}]^T}}} }) \in {{\mathbb{C}}^{N \times N}}$ is the reflection coefficient matrix of the $r$-th RIS; ${\varphi_{r,n}}$ is the phase shift imposed by the $n$-th element of the $r$-th RIS; ${\boldsymbol{\theta}} \buildrel \Delta \over = {e^{\left( {j{{\left[ {{\varphi _{1,1}}, \cdots ,{\varphi _{1,N}},{\varphi _{2,1}}, \cdots ,{\varphi _{R,N}}} \right]}^T}} \right)}} \in {{\mathbb{C}}^{NR \times 1}}$ denotes the phase shift vector of $R$ RISs;  ${{\bf{V}}_k} \buildrel \Delta \over = {\rm{diag}}( {[{\bf{v}}_{1,k}^T,{\bf{v}}_{2,k}^T,}$ ${\cdots ,{\bf{v}}_{R,k}^T]} )$ $\in {{\mathbb{C}}^{NR \times NR}}$ represents the equivalent channel from $R$ RISs to the $k$-th UE; ${{\bf{G}}_b} = {[{\bf{G}}_{b,1}^T,{\bf{G}}_{b,2}^T, \cdots ,}$ ${{\bf{G}}_{b,R}^T]^T} \in {{\mathbb{C}}^{NR \times {N_t}}}$ denotes the equivalent channel from the $b$-th BS to $R$ RISs; ${\bf{\hat h}}_{b,k}$ is the composite channel from the $b$-th BS to the $k$-th UE, incorporating one direct and $NR$ reflected channels.

We assume that all BSs are synchronized to ensure joint service for all UEs in the same time-frequency resource block. Therefore, the received signal at the $k$-th UE is the superposition of the signals transmitted from all  BSs, which can be expressed as  
\begin{align}   \label{eq5}
{y_k} &= \sum\limits_{b = 1}^B {{y_{b,k}} + {z_k}}\notag\\
&= \sum\limits_{b = 1}^B {\sum\limits_{j = 1}^K {{\bf{\hat h}}_{b,k}^H} } {{\bf{w}}_{b,j}}{s_j} + {z_k}\notag\\
&= \underbrace {\sum\limits_{b = 1}^B {{\bf{\hat h}}_{b,k}^H} {{\bf{w}}_{b,k}}{s_k}}_{{\rm{Desired}}\;{\rm{signal}}} + \underbrace {\sum\limits_{b = 1}^B {\sum\limits_{j = 1,j \ne k}^K {{\bf{\hat h}}_{b,k}^H} } {{\bf{w}}_{b,j}}{s_j}}_{{\rm{Interference}}\;{\rm{of}}\;{\rm{other}}\;{\rm{UEs}}} + {z_k},
\end{align}
where ${z_k}\sim {\cal C}{\cal N}(0,\delta _k^2)$ denotes the additive white Gaussian noise (AWGN). Without loss of generality, we assume that all UEs have the same noise power, i.e., $\delta _k^2 = {\delta ^2},\forall k \in \cal K$.


\subsection{Problem Formulation}
Based on the signal model expressed in (\ref{eq5}), the signal-to-interference-plus-noise ratio (SINR) ${\zeta _k}$ of the $k$-th UE can be written as
\begin{align}   \label{eq6}
{\zeta _k} = \frac{{{{\left| {\sum\limits_{b = 1}^B {{\bf{\hat h}}_{b,k}^H} {{\bf{w}}_{b,k}}} \right|}^2}}}{{\sum\limits_{j = 1,j \ne k}^K {{{\left| {\sum\limits_{b = 1}^B {{\bf{\hat h}}_{b,k}^H} {{\bf{w}}_{b,j}}} \right|}^2} + {\delta ^2}}}}.
\end{align}

To evaluate the performance of the RIS-assisted cell-free system, the WSR is given as
\begin{align}   \label{eq7}
{\rm{WSR}} = \sum\limits_{k = 1}^K {{\omega _k}{{\log }_2}\left( {1 + {\zeta _k}} \right)},
\end{align}
where ${\omega _k}>0$ is the weight of the $k$-th UE, which indicates the priority of different UEs. 

In this paper, we endeavor to maximum the WSR of the RIS-assisted cell-free system by designing the BS precoding ${\bf{W}} = \left\{ {{{\bf{w}}_{b,k}}|{\forall b \in \cal B},{\forall k \in \cal K}} \right\}$ and the RIS reflection coefficient vector ${\boldsymbol{\theta}}$. Mathematically, the optimization problem can be formulated as 
\begin{subequations} \label{eq8}
\begin{align}   
	{\cal P}0:\mathop {\;\;\max }\limits_{{\boldsymbol{\theta}},{\bf{W}}} \;\;&{\rm{WSR}} = \sum\limits_{k = 1}^K {{\omega _k}{{\log }_2}\left( {1 + {\zeta _k}} \right)}\label{eq8a}\\
	s.t.\;\;
	&\sum\limits_{k = 1}^K {{{\left\| {{{\bf{w}}_{b,k}}} \right\|}^2}}  \le {P_{b,\max }},\;\forall b \in {\cal B},\label{eq8b}\\
	&\left| {{\theta _{r, b}}} \right| = 1,\;\forall r \in {\cal R},\;\forall n \in {\cal N},\label{eq8c}
\end{align}
\end{subequations}
where (\ref{eq8a}) is the WSR objective function; (\ref{eq8b}) is the power constraints of BSs, where ${P_{b,\max }}$ denotes the maximum transmit power budget at the $b$-th BS. Constraint (\ref{eq8c}) represents that the amplitude of reflection coefficient of each RIS remains constant in this paper. 

\emph{Remark 1:} Although the centralized algorithm can achieve the optimal solution to ${\cal P}0$ \cite{RIScellfree4}, it requires collecting the local CSI of all BSs for joint optimization at the CPU. This inevitably increases both the CSI feedback and control signaling overhead as well as the computational complexity of the CPU. Therefore, we aim to develop a distributed algorithm to solve ${\cal P}0$ by spreading the computational load to the distributed BSs.

Therefore, the distributed optimation problem of ${\cal P}0$ is rewritten as
\begin{subequations} \label{eq9}
\begin{align}   
	{\cal P}1:\mathop {\;\;\max }\limits_{{\boldsymbol{\theta}},{\bf{W}}} \;\;&{\rm{WSR}} = \sum\limits_{k = 1}^K {{\omega _k}{{\log }_2}\left( {1 + {\zeta _k}} \right)}\label{eq9a}\\
	s.t.\;\;
	&{(9b), (9c),}\notag\\	
	&{{\boldsymbol{\theta}} _b} = {{\boldsymbol{\theta}} _{\bar b}},\forall b \in {\cal B},\forall {\bar b} \in {\cal F}_b,\label{eq9b}
\end{align}
\end{subequations}
where (\ref{eq9b}) is the consensus constraint, which means that ${{\boldsymbol{\theta}} _b}$ optimized at adjacent BSs should be consistent. ${\cal F}_b$ represents the index set of the adjacent BSs that can exchange information with the $b$-th BS. Specifically, the $b$-th BS requires utilizing the information from the adjacent BSs when designing the RIS reflection coefficients. Then, it sends its local information to the adjacent BSs until the RIS reflection coefficients on all BSs reach a consensus.

\emph{Remark 2:} Here we highlight that the optimization of ${\bf{W}}_b$ and $\boldsymbol{\theta }_b$ in a distributed system are distinctly different. Specifically, the downlink precoding matrix ${\bf{W}}_b$ is unique for different BSs. By contrast, $\boldsymbol{\theta }_b$ optimized by different BSs correspond to the same RIS and need to be appropriately fused into a single reflection coefficient vector, which is known as the \textit{consensus} problem in distributed systems. Although the centralized algorithm does not involve the consensus problem, the distributed optimization strategy is more effective and practical considering the distributed deployment of BSs as well as the limited backhaul capacity.

\section{ADMM-Based Distributed Optimization}
In this section, we propose a distributed ADMM-based design for effectively optimizing the precoder and reflection phase shifts in the practical RIS-assisted cell-free system. Specifically, we first convert the non-convex ${\cal P}1$ into a tractable form ${\cal P}2$. Then, we propose a distributed ADMM design to solve ${\cal P}2$.


\subsection{A Tractable Form of ${\cal P}1$}
Observe form (\ref{eq9a}) that ${\cal P}1$ is a non-convex optimization problem due to the coupling of the optimization variables ${\bf{W}}$ and ${\boldsymbol{\theta}}$ and the consensus constraint (\ref{eq9b}). Therefore, we transform ${\cal P}1$ into a tractable problem by applying the Lagrangian dual transform and the  quadratic transform, which are summarized in Lemmas 1 and 2, respectively.

\textit{Lemma 1 (Lagrangian dual transform):} Given a sum-of-logarithmic-ratios problem, expressed as {\cite{LT2}}
\begin{subequations} \label{eq10}
\begin{align}   
	\mathop {\;\;\max }\limits_{\bf x}  \;\;&\sum\limits_{d = 1}^D {{\omega _d}{{\log }_2}\left( {1 + \frac{{{Q_d}\left( {\bf x}  \right)}}{{{F_d}\left( {\bf x}  \right)}}} \right)}\label{eq10a}\\
	s.t.\;\;
	&{\bf{x}} \in \chi ,\label{eq10b}			
\end{align}
\end{subequations}
where ${\omega _d}$ is a nonnegative weight; ${{Q_d}\left( {\bf{x}} \right)}$ is a nonnegative function that satisfies ${Q_d}\left( {\bf{x}} \right) \ge 0$; ${{F_d}\left( {\bf{x}} \right)}$ is a positive function with ${F_d}\left( {\bf{x}} \right) > 0$; ${\bf x}$ is the optimization variable, and $\chi$ denotes a nonempty constraint set. Moving the ratio from inside of the logarithm to the outside, (\ref{eq10a}) can be rewritten as  
\begin{subequations} \label{eq11}
\begin{align}   
	\mathop {\;\;\min }\limits_{{\bf{x}},{\boldsymbol{\gamma}} } \;\;&\sum\limits_{d = 1}^D {{\omega _d}\left( {{\gamma _d} - {{\log }_2}\left( {1 + {\gamma _d}} \right) - \frac{{\left( {1 + {\gamma _d}} \right){Q_d}\left( {\bf{x}} \right)}}{{{Q_d}\left( {\bf{x}} \right) + {F_d}\left( {\bf{x}} \right)}}} \right)} \label{eq11a}\\
	s.t.\;\;
	&{\bf{x}} \in \chi , {\gamma _d} \le \frac{{{Q_d}({\bf{x}})}}{{{F_d}({\bf{x}})}},\label{eq11b}			
\end{align}
\end{subequations}
where ${\boldsymbol{\gamma }} = [{\gamma _1},{\gamma _2}, \cdots {\gamma _D}]^T$ is the auxiliary variable vector.

\textit{Lemma 2 (Quadratic transform):} Given a sum-of-functions-of-ratio problem for the multidimensional and complex cases, expressed as {\cite{LT1}}
\begin{subequations} 
	\begin{align}   
		\mathop {\;\;\min }\limits_{\bf x}  \;\;&\sum\limits_{d = 1}^D {{{\bar f}_d}\left( {{Q_d}\left( {\bf{x}} \right)F_d^{ - 1}\left( {\bf{x}} \right)Q_d^H\left( {\bf{x}} \right)} \right)}  \label{eq14a}\\
		s.t.\;\;
		&{\bf{x}} \in \chi ,\label{eq14b}			
	\end{align}
\end{subequations}
where function ${Q_d}\left( {\bf{x}} \right): {{\mathbb{C}}^{{d_1}}} \to {{\mathbb{C}}^{{d_2}}}$, ${F_d}\left( {\bf{x}} \right) {{\mathbb{C}}^{{d_1}}} \to {{\mathbb{S}}^{{d_2} \times {d_2}}}$, ${\bf x}$, and constraint $\chi  \subseteq {{\mathbb{C}}^{{d_1}}}$. Let ${{\bar f}_d}\left(  \cdot  \right)$ denotes a monotonically nondecreasing function, problem (13) can be transformed to
\begin{subequations} \label{eq15}
	\begin{align}   
		\mathop {\;\;\min }\limits_{{\bf{x}},{\boldsymbol{\eta}} } \;\;&\sum\limits_{d = 1}^D {{{\bar f}_d}\left( {2{\mathop{\rm Re}\nolimits} \left\{ {{\eta _d}{Q_d}\left( {\bf{x}} \right)} \right\} - \eta _d^2{F_d}\left( {\bf{x}} \right)} \right)}  \label{eq15a}\\
		s.t.\;\;
		&{\bf{x}} \in \chi , {\eta _d} \in {{\mathbb{C}}^{{d_1}}},\label{eq15b}			
	\end{align}
\end{subequations}
where ${\boldsymbol{\eta }} = [{\eta _1},{\eta _2}, \cdots {\eta _D}]^T$ denotes the auxiliary variable vector.

Therefore, by using the Lagrangian dual transform, ${\cal P} 1$ can be reformulated as 
\begin{subequations} \label{eq12}
\begin{align}   
	{\cal L}{\cal P}1:\mathop {\;\;\min }\limits_{{\boldsymbol{\theta }},{\bf{W}},{\boldsymbol{\gamma }}} \;\;&{f_1}\left( {{\boldsymbol{\theta }},{\bf{W}},{\boldsymbol{\gamma }}} \right)\\
	s.t.\;\;
	&{(9b), (9c), (10b)},\notag		
\end{align}
\end{subequations}
where ${f_1}\left( {{\boldsymbol{\theta }},{\bf{W}},{\boldsymbol{\gamma }}} \right)$ is the new objective function via Lemma 1, which is described in (\ref{eq13}). Besides, ${\boldsymbol{\gamma }} = [{\gamma _1},{\gamma _2}, \cdots {\gamma _K}]^T$ represents the auxiliary variable vector. 
\begin{align}   \label{eq13}
	{f_1}\left( {{\boldsymbol{\theta }},{\bf{W}},{\boldsymbol{\gamma }}} \right) = \sum\limits_{k = 1}^K {{\omega _k}\left( {{\gamma _k} - {{\log }_2}\left( {1 + {\gamma _k}} \right) - \frac{{\left( {1 + {\gamma _k}} \right){{\left| {\sum\limits_{b = 1}^B {{\bf{\hat h}}_{b,k}^H} {{\bf{w}}_{b,k}}} \right|}^2}}}{{\sum\limits_{k = 1}^K {{{\left| {\sum\limits_{b = 1}^B {{\bf{\hat h}}_{b,k}^H} {{\bf{w}}_{b,k}}} \right|}^2} + {\delta ^2}} }}} \right)}. 
\end{align}

Then we use the quadratic transform shown in Lemma 2 to decouple the numerator and the denominator of the fraction in ${\cal L}{\cal P}1$ to further simplify the optimization. Consequently, ${\cal L}{\cal P}1$ can be transformed as
\begin{subequations} \label{eq16}
\begin{align}   
	{\cal L}{\cal P}2:\mathop {\;\;\min }\limits_{{\boldsymbol{\theta }},{\bf{W}},{\boldsymbol{\gamma }},{\boldsymbol{\eta}}} \;\;&{f_2}\left( {{\boldsymbol{\theta }},{\bf{W}},{\boldsymbol{\gamma }},{\boldsymbol{\eta}} } \right)\label{eq16a}\\
	s.t.\;\;
	&{(9b), (9c), (10b)},\notag
\end{align}
\end{subequations}
where ${f_2}\left( {{\boldsymbol{\theta }},{\bf{W}},{\boldsymbol{\gamma }},{\boldsymbol{\eta}} } \right)$ is given in (\ref{eq17}); ${\boldsymbol{\eta }} = [{\eta _1},{\eta _2}, \cdots {\eta _K}]^T$ denotes the auxiliary variable vector.
\begin{align}   \label{eq17}
	{f_2}\left( {{\bf{\theta }},{\bf{W}},{\bf{\gamma }},{\bf{\eta }}} \right) &= \sum\limits_{k = 1}^K {({{\left| {{\eta _k}} \right|}^2}\sum\limits_{j = 1}^K {{{\left| {\sum\limits_{b = 1}^B {{\bf{\hat h}}_{b,j}^H} {{\bf{w}}_{b,k}}} \right|}^2}} }  + {\left| {{\eta _k}} \right|^2}{\delta ^2} + {\omega _k}{\gamma _k}\notag\\
	& - 2\sqrt {\left( {1 + {\gamma _k}} \right){\omega _k}} \sum\limits_{b = 1}^B {{\mathop{\rm Re}\nolimits} } \left\{ {{\eta _k}{\bf{\hat h}}_{b,k}^H{{\bf{w}}_{b,k}}} \right\} 
	 - {\omega _k}{\log _2}\left( {1 + {\gamma _k}} \right). 
\end{align}

Next, we rewrite problem ${\cal L}{\cal P}2$ in its augmented Lagrangian form, expressed as
\begin{align}   \label{eq18}
	{\cal L}\left( {{\boldsymbol{\theta }},{\bf{W}},{\boldsymbol{\gamma }},{\boldsymbol{\eta}} ,{\boldsymbol{\lambda}} } \right) &= {f_2}\left( {{\boldsymbol{\theta }},{\bf{W}},{\boldsymbol{\gamma }},{\boldsymbol{\eta}} } \right) + \underbrace {\sum\limits_{b = 1}^B {{\mu _b}\left( {\sum\limits_{k = 1}^K {{{\left\| {{{\bf{w}}_{b,k}}} \right\|}^2}}  - {P_{b,\max }}} \right)} }_{{\rm{Power}}\;{\rm{constraint}}}\notag\\
	& + \underbrace {\sum\limits_{b = 1}^B {{\mathbb{I}}\left( {{{\boldsymbol{\theta}} _b}} \right)} }_{{\rm{Feasible}}\;{\rm{constraint}}}  
	+ \underbrace {{{\sum\limits_{b = 1}^B {\frac{{{\rho _b}}}{2}\left\| {{{\boldsymbol{\theta}} _b} - {{\boldsymbol{\theta}} _{\bar b}} + \frac{{{{\boldsymbol{\lambda}} _b}}}{{{\rho _b}}}} \right\|} }^2}}_{{\rm{Consensus}}\;{\rm{constraint}}}.
\end{align}
where ${\boldsymbol{\lambda }} = \left\{ {{\boldsymbol{\lambda } _b} \in {\mathbb{C}^{N \times 1}}|\forall b \in B} \right\}$ is the Lagrange multiplier and $\rho _{b} > 0$. ${\mathbb{I}}\left(  \cdot  \right)$ represents a feasible function such that ${{\mathbb{I}}\left( {{{\boldsymbol{\theta}} _b}} \right)}=0$ for $\left| {{\theta _b}} \right| = 1$ and ${{\mathbb{I}}\left( {{{\boldsymbol{\theta}} _b}} \right)}=\infty $ for $\left| {{\theta _b}} \right| \ne 1$. As a result, the final tractable form of ${\cal P}1$ is given as 
\begin{align}   \label{eq19}
{\cal P}2:\mathop {\;\;\min }\limits_{{\boldsymbol{\theta }},{\bf{W}},{\boldsymbol{\gamma }},{\boldsymbol{\eta}}, {\boldsymbol{\lambda}}} {\cal L}\left( {{\boldsymbol{\theta }},{\bf{W}},{\boldsymbol{\gamma }},{\boldsymbol{\eta}} ,{\boldsymbol{\lambda}} } \right)		
\end{align}

\subsection{Proposed Distributed Design Based on ADMM}
To solve the problem ${\cal P}2$, we propose a distributed design based on ADMM \cite{ADMM1, ADMM2}. The proposed design iteratively designs the local BS precoding and RIS reflection coefficient vectors at each BS. Specifically, the $i$-th iteration at the $b$-th BS can be expressed as  
\begin{subequations} \label{eq20}
\begin{align}   
	&{{\boldsymbol{\gamma }}^i} = \arg \;\mathop {\min }\limits_{\boldsymbol{\gamma }} {f_1}\left( {{{\boldsymbol{\theta }}^{i - 1}},{{\bf{W}}^{i - 1}},{\boldsymbol{\gamma }}} \right),\label{eq20a}\\
	&{{\boldsymbol{\eta }}^i} = \arg \;\mathop {\min }\limits_{\boldsymbol{\eta }} {\cal L}\left( {{{\boldsymbol{\theta }}^{i - 1}},{{\bf{W}}^{i - 1}},{\boldsymbol{\gamma }^{i}},{{\boldsymbol{\eta }}},{{\boldsymbol{\lambda }}^{i - 1}}} \right),\label{eq20b}\\
	&{\bf{W}}_b^i = \arg \;\mathop {\min }\limits_{\boldsymbol{\bf{W}}_b } {{\cal L}}\left( {{{\boldsymbol{\theta }}^{i - 1}},{\bf{\bar W}}_b^{i - 1},{{\bf{W}}_b},{\boldsymbol{\gamma }^{i}},{{\boldsymbol{\eta }}^{i}},{{\boldsymbol{\lambda }}^{i - 1}}} \right),\label{eq20c}\\
	&{\boldsymbol{\theta }}_b^i = \arg \;\mathop {\min }\limits_{{{\boldsymbol{\theta }}_b}} {\cal L}\left( {{\boldsymbol{\bar \theta }}_b^{i - 1},{{\boldsymbol{\theta }}_b},{\bf{\bar W}}_b^{i - 1},{\bf{W}}_b^i,{{\boldsymbol{\gamma }}^i},{{\boldsymbol{\eta }}^i},{{\boldsymbol{\lambda }}^{i - 1}}} \right),\label{eq20d}\\
	&{\boldsymbol{\lambda }}_b^i = {\boldsymbol{\lambda }}_b^{i - 1} + {\rho _b}\left( {{\boldsymbol{\theta }}_b^i - {\boldsymbol{\theta }}_{_{\bar b}}^i} \right),\label{eq20e}
\end{align}
\end{subequations}
where $\;{{\bf{W}}_b} = \left\{ {{{\bf{w}}_{b,k}}|\forall k \in \cal K} \right\}$ denotes the local precoding at the $b$-th BS; ${{{\bf{\bar W}}}_b} = {\bf{W}}\backslash {{\bf{W}}_b}$ and ${{{\boldsymbol{\bar \theta}}}_b} = {\boldsymbol{\theta }}\backslash {{\boldsymbol{\theta }}_b}$ are the downlink precoding and the RIS reflection coefficient vector of other BSs except the $b$-th BS; Observe from (\ref{eq20}) that $\boldsymbol{\gamma }$, $\boldsymbol{\eta }$, ${\bf{W}}_b$, $\boldsymbol{\theta }_b$, and $\boldsymbol{\lambda }_b$ are updated locally in sequence.

Next, we give the solutions to problems (\ref{eq20a})-(\ref{eq20d}) one by one. Note that ${\cal P}2$ is an equivalence problem to ${\cal L}{\cal P}1$, which means that the optimal $\gamma$ of ${\cal L}{\cal P}1$ is equal to ones of ${\cal P}2$. Besides, it is easier to solve ${\cal L}{\cal P}1$ than ${\cal P}2$ for optimal $\gamma$. Therefore, we solve ${\cal L}{\cal P}1$ for optimal $\gamma$.

\subsubsection{The Solver of (\ref{eq20a})}
For problem (\ref{eq20a}), ${f_1}\left( {{\boldsymbol{\theta }},{\bf{W}},{\boldsymbol{\gamma }}} \right)$ is a convex function for $\boldsymbol{\gamma }$ with fixed $\boldsymbol{\theta }$ and $\bf{W}$. Therefore, the optimal ${\gamma }_k$ can be obtained by taking $\frac{{\partial \left( {f_1}\left( {{\boldsymbol{\gamma }} } \right) \right)}}{{\partial {{\bf{\gamma }}_k}}} = 0$. Thus, we have  
\begin{align}   \label{eq21}
\gamma _k^\dag  = \frac{{{{\left| {\sum\limits_{b = 1}^B {{\bf{\hat h}}_{b,k}^H} {{\bf{w}}_{b,k}}} \right|}^2}}}{{\sum\limits_{j = 1,j \ne k}^K {{{\left| {\sum\limits_{b = 1}^B {{\bf{\hat h}}_{b,k}^H} {{\bf{w}}_{b,j}}} \right|}^2} + {\delta ^2}} }}= \frac{{{{\left| {{\varpi _{k,k}} + {\vartheta _{k,k}}} \right|}^2}}}{{\sum\limits_{j = 1,j \ne k}^K {{{\left| {{\varpi _{k,j}} + {\vartheta _{k,j}}} \right|}^2} + {\delta ^2}} }},		
\end{align}
where ${\varpi _{k,j}} = \sum\limits_{b = 1}^B {{\bf{h}}_{b,k}^H{{\bf{w}}_{b,j}}}$ and ${\vartheta _{k,j}} = \sum\limits_{b = 1}^B {{\boldsymbol{\theta} ^H}{\bf{V}}_k^H{{\bf{G}}_b}{{\bf{w}}_{b,j}}}$ are two defined cross-term information, which contain the information of all BSs. Note that $\left\{ {{\varpi _{k,j}}|\forall k,j \in \cal K} \right\}$ and $\left\{ {{\vartheta _{k,j}}|\forall k,j \in \cal K} \right\}$ are then exchanged among different BSs to achieve the goal of cooperative design.

\subsubsection{The Solver of (\ref{eq20b})}
For problem (\ref{eq20b}), we note that only ${f_2}\left( {{\boldsymbol{\theta }},{\bf{W}},{\boldsymbol{\gamma }},{\boldsymbol{\eta}} } \right)$ in (\ref{eq18}) is dependent on $\boldsymbol{\eta }$. Therefore, problem (\ref{eq20b}) can be reformulated as 
\begin{align}   \label{eq22}
{{\boldsymbol{\eta }}} &= \arg \;\mathop {\min }\limits_{\boldsymbol{\eta }} \sum\limits_{k = 1}^K {\left( {{{\left| {{\eta _k}} \right|}^2}\left( {\sum\limits_{j = 1}^K {{{\left| {\sum\limits_{b = 1}^B {{\bf{\hat h}}_{b,j}^H} {{\bf{w}}_{b,k}}} \right|}^2} + {\delta ^2}} } \right)} \right.}\notag \\
&\left. { - \sqrt {\left( {1 + {\gamma _k}} \right){\omega _k}} \left( {\sum\limits_{b = 1}^B {\left( {{\eta _k}{\bf{\hat h}}_{b,k}^H{{\bf{w}}_{b,k}} + \eta _k^ * {\bf{w}}_{b,k}^H{{{\bf{\hat h}}}_{b,k}}} \right)} } \right)} \right).		
\end{align}

The optimal ${{\eta }} _k^\dag $ can also be obtained by taking $\frac{{\partial \left( {f_2}\left( \boldsymbol{\eta }  \right) \right)}}{{\partial {{\bf{\eta }}_k^*}}} = 0$, which is expressed as
\begin{align}   \label{eq23}
\eta _k^\dag  = \frac{{{{\left( {{\varpi _{k,k}} + {\vartheta _{k,k}}} \right)}^ * }\sqrt {\left( {1 + {\gamma _k}} \right){\omega _k}} }}{{\sum\limits_{j = 1}^K {{{\left| {{\varpi _{k,j}} + {\vartheta _{k,j}}} \right|}^2} + {\delta ^2}} }}.		
\end{align}

\subsubsection{The Solver of (\ref{eq20c})}
Given a set of tentative values of other variables, we have 
\begin{align}   \label{eq24}
	{\cal L}\left( {{{\bf{w}}_{b,k}}} \right) &= \sum\limits_{k = 1}^K ( {{{\left| {{\eta _k}} \right|}^2}\sum\limits_{j = 1}^K {\left( {{{\left\| {\left( {\sum\limits_{{b^{'}} \ne b}^B {{\bf{\hat h}}_{{b^{'}},j}^H{{\bf{w}}_{{b^{'}},k}}} } \right){\bf{w}}_{b,k}^H{{{\bf{\hat h}}}_{b,j}}} \right\|}^2} + {{\left\| {{\bf{h}}_{b,j}^H{{\bf{w}}_{b,k}}} \right\|}^2}} \right)}}   \notag\\
	&- \sqrt {\left( {1 + {\gamma _k}} \right){\omega _k}} {\mathop{\rm Re}\nolimits} \left\{ {{\eta _k}{\bf{\hat h}}_{b,k}^H{{\bf{w}}_{b,k}}} \right\} ) +{\mu _b}{\left\| {{{\bf{w}}_{b,k}}} \right\|^2} + {{\bf{C}}_1}.
\end{align}
where ${{\bf{C}}_1}$ is defined by
\begin{align}   \label{eq25}
	{{\bf{C}}_1} &= \sum\limits_{k = 1}^K {( {{{\left| {{\eta _k}} \right|}^2}\sum\limits_{j = 1}^K {\sum\limits_{{b^{'}} \ne b}^B {{{\left\| {{\bf{\hat h}}_{{b^{'}},j}^H{{\bf{w}}_{{b^{'}},k}}} \right\|}^2}\sum\limits_{j = 1}^K { - 2\sqrt {\left( {1 + {\gamma _k}} \right){\omega _k}} \sum\limits_{{b^{'}} \ne b}^B {{\mathop{\rm Re}\nolimits} \left\{ {{\eta _k}{\bf{\hat h}}_{{b^{'}},k}^H{{\bf{w}}_{{b^{'}},k}}} \right\}} }  + } } {{\left| {{\eta _k}} \right|}^2}{\delta ^2}}}+ {\omega _k}{\gamma _k}\notag\\
	& - {{{\omega _k}{{\log }_2}\left( {1 + {\gamma _k}} \right)} )} + \sum\limits_{{b^{'}} \ne b}^B {{\mu _{{b^{'}}}}\left( {\sum\limits_{k = 1}^K {{{\left\| {{{\bf{w}}_{{b^{'}},k}}} \right\|}^2}}  - {P_{b,\max }}} \right)}  + {\mu _b}\left( {\sum\limits_{{k^{'}} \ne k}^K {{{\left\| {{{\bf{w}}_{b,{k^{'}}}}} \right\|}^2}}  - {P_{b,\max }}} \right) \notag\\
	&+ \sum\limits_{b = 1}^B {\mathbb{I}\left( {{{\boldsymbol{\theta }}_b}} \right)}  + {\sum\limits_{b = 1}^B {\frac{{{\rho _b}}}{2}\left\| {{{\boldsymbol{\theta }}_b} - {{\boldsymbol{\theta }}_{\bar b}} + \frac{{{{\boldsymbol{\lambda }}_b}}}{{{\rho _b}}}} \right\|} ^2}. 
\end{align}

Similarly, the optimal ${\bf{w}}_{b,k}^\dag $ can be obtained by taking $\frac{{\partial \left( {{\cal L}\left( {{{\bf{w}}_{b,k}}} \right)} \right)}}{{\partial {\bf{w}}_{b,k}^H}} = 0$, which is given as 
\begin{align}   \label{eq26}
{\bf{w}}_{b,k}^\dag  = \frac{{\sqrt {\left( {1 + {\gamma _k}} \right){\omega _k}} \eta _k^ * {{{\bf{\hat h}}}_{b,k}} - {\boldsymbol{\Omega} _{b,k}}}}{{\left( {{\mu _b} + {{{\left| {{\eta _k}} \right|}^2}} \sum\limits_{j = 1}^K  {{{\bf{\hat h}}}_{b,j}}{\bf{\hat h}}_{b,j}^H} \right)}},		
\end{align}
where ${\boldsymbol{\Omega} _{b,k}} = {{\left| {{\eta _k}} \right|}^2} \sum\limits_{j = 1}^K {{{{\bf{\hat h}}}_{b,k}}( {{\varpi _{j,k}} + {\vartheta _{j,k}} - {\bf{\hat h}}_{b,j}^H{{\bf{w}}_{b,k}}} )}$; ${{\mu _b}}$ is a normalized factor used to scale ${{{\bf{w}}_{b,k}}}$ for satisfying the total power constraint, which needs to be  dynamically updated in each iteration. We apply the power normalization approach below to bypass the update of ${\mu _b}$ in D$^2$-ADMM. Specifically, ${\bf{w}}_{b,k}^l$ is scaled by
\begin{align}   \label{eq34}
	{\bf{w}}_{b,k}^{{\rm{\dag}}} = \frac{{\sqrt {{P_{b,\max }}} {\bf{w}}_{b,k}^{{\rm{\dag}}}}}{{\sqrt {\sum\limits_{k = 1}^K  {\left\| {{{\bf{w}}_{b,k}^{\rm{\dag}}}} \right\|}^2} }}. 
\end{align}

\subsubsection{The Solver of (\ref{eq20d})}
For problem (\ref{eq20d}), we first rewrite (\ref{eq18}) in a more intuitive form as 
\begin{align}   \label{eq27}
{\cal L}\left( {{{\boldsymbol{\theta }}_b}} \right) = {\boldsymbol{\theta }}_b^H{\bf{S}}{{\boldsymbol{\theta }}_b} - 2{\mathop{\rm Re}\nolimits} \left\{ {{\boldsymbol{\theta }}_b^H{\bf{Z}}} \right\} + {\sum\limits_{b = 1}^B {{\mathbb{I}}\left( {{{\boldsymbol{\theta}} _b}} \right)} }+ {{\bf{C}}_2},		
\end{align}
where ${\bf{S}}$, ${\bf{Z}}$, and ${{\bf{C}}_2}$ are independent of ${{{\boldsymbol{\theta }}_b}}$ and given in (\ref{eq28}), (\ref{eq29}), and (\ref{eq30}) respectively. Note that (\ref{eq27}) is a non-convex problem due to the feasible constraint. To solve this problem, we build a neural block based on DL, and the details will be discussed in Section IV.
\begin{align}   \label{eq28}
	{\bf{S}} = \sum\limits_{k = 1}^K {{\left| {{\eta _j}} \right|}^2} {\sum\limits_{j = 1}^K {{\bf{V}}_j^H{{\bf{G}}_b}{{\bf{w}}_{b,k}}} } {\bf{w}}_{b,k}^H{\bf{G}}_b^H{{\bf{V}}_j} + \frac{{{\rho _b}}}{2}.
\end{align}
\begin{align}   \label{eq29}
	{\bf{Z}} = \sum\limits_{k = 1}^K {{\left| {{\eta _k}} \right|}^2} {\sum\limits_{j = 1}^K {{\bf{V}}_j^H{{\bf{G}}_b}{{\bf{w}}_{b,k}}} } \left( {{\bf{w}}_{b,k}^H{{{\bf{\hat h}}}_{b,j}} - \varpi _{j,k}^ *  - \vartheta _{j,k}^ *  - {\bf{w}}_{b,k}^H{{\bf{h}}_{b,j}}} \right) + \sqrt {\left( {1 + {\gamma _k}} \right){\omega _k}} {\eta _k}{\bf{V}}_k^H{{\bf{G}}_b}{{\bf{w}}_{b,k}}.
\end{align}
\begin{align}   \label{eq30}
	{{\bf{C}}_2} &= \sum\limits_{k = 1}^K {\left( {{\left| {{\eta _k}} \right|}^2}{\sum\limits_{j = 1}^K {\left( {\sum\limits_{{b^{'}} \ne b}^B {{{\left\| {{\bf{\hat h}}_{{b^{'}},j}^H{{\bf{w}}_{{b^{'}},k}}} \right\|}^2}}  + 2{\mathop{\rm Re}\nolimits} \left\{ {{\bf{h}}_{b,j}^H{{\bf{w}}_{b,k}}\left( {\sum\limits_{{b^{'}} \ne b}^B {{\bf{w}}_{{b^{'}},k}^H{{{\bf{\hat h}}}_{{b^{'}},j}}} } \right)} \right\} + {{\left\| {{\bf{h}}_{b,j}^H{{\bf{w}}_{b,k}}} \right\|}^2}} \right)} } \right.} \notag\\
	&- \left. {\sqrt {\left( {1 + {\gamma _k}} \right){\omega _k}} \sum\limits_{{b^{'}} \ne b}^B {{\mathop{\rm Re}\nolimits} \left\{ {{\eta _k}{\bf{\hat h}}_{{b^{'}},k}^H{{\bf{w}}_{{b^{'}},k}}} \right\}}  + {\omega _k}{\gamma _k} - {\omega _k}{{\log }_2}\left( {1 + {\gamma _k}} \right) + {{\left| {{\eta _k}} \right|}^2}{\delta ^2}} \right)   \notag\\
	&  +    		\sum\limits_{{b^{'}} \ne b}^B {\frac{{{\rho _{{b^{'}}}}}}{2}{{\left\| {{{\boldsymbol{\theta }}_{{b^{'}}}} - {{\boldsymbol{\theta }}_{{{\bar b}^{'}}}} + \frac{{{{\boldsymbol{\lambda }}_{{b^{'}}}}}}{{{\rho _{{b^{'}}}}}}} \right\|}^2} + } \frac{{{\rho _b}}}{2}{\left\| {\frac{{{{\boldsymbol{\lambda }}_b}}}{{{\rho _b}}} - {{\boldsymbol{\theta }}_{\bar b}}} \right\|^2}.
\end{align}

%
%

\section{D$^2$-ADMM: A Learning-based algorithm unrolling Method}
This section proposes a D$^2$-ADMM neural network structure to design the BS precoding and the RIS reflection coefficient by unfolding the proposed distributed ADMM design. Furthermore, an efficient monodirectional information exchange strategy is proposed to link different BSs to improve the performance of our distributed designs. Finally, we elaborate on the training and the implementation of D$^2$-ADMM. 

\subsection{Structure of Deep Distributed ADMM}
\begin{figure*}[tbp]
\centering {
	\begin{tabular}{ccc}
		\includegraphics[width=1\textwidth]{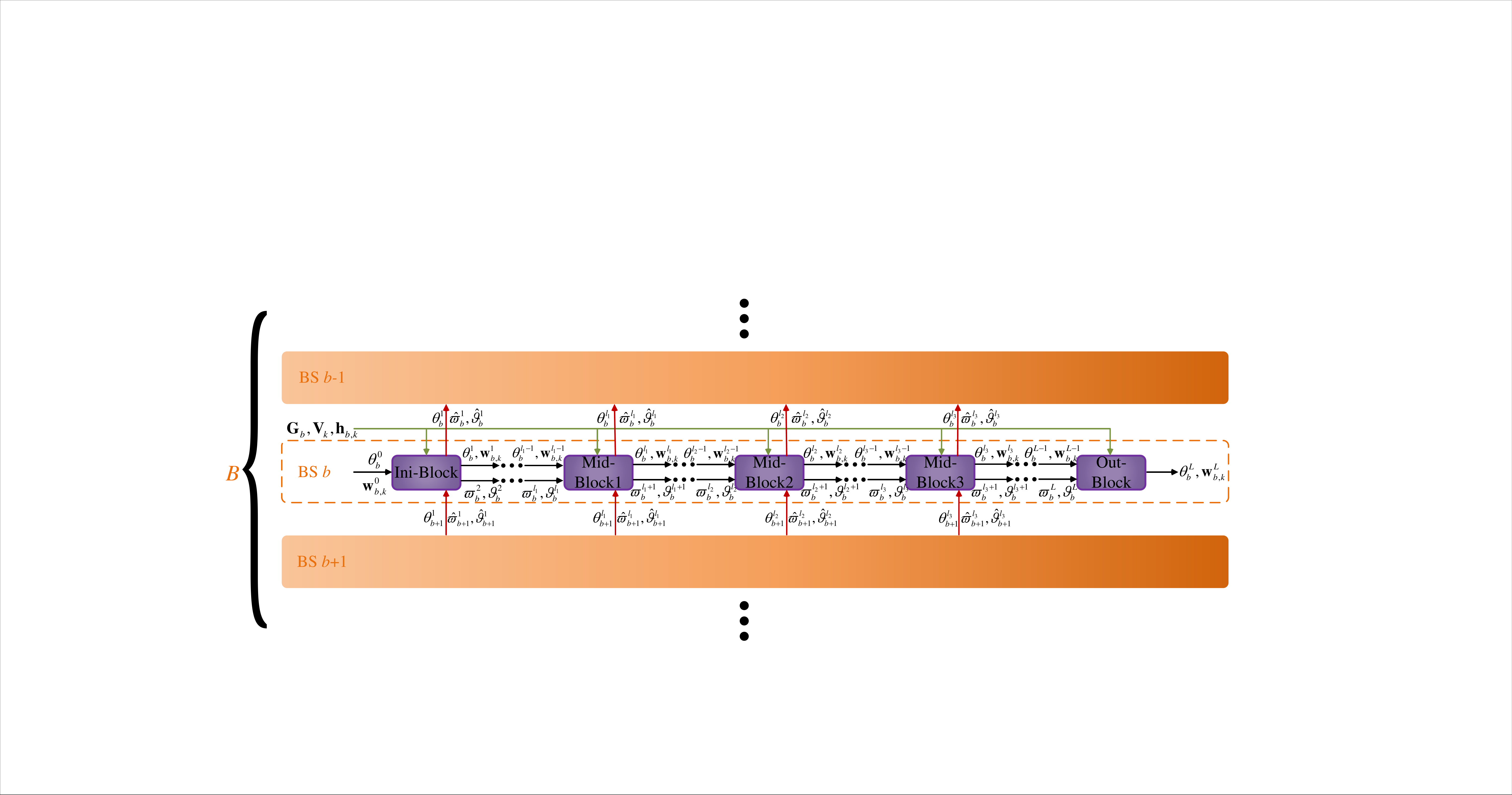}\\
	\end{tabular}
}
\caption{The structure of D$^2$-ADMM.}
\vspace{-1\baselineskip}
\label{deepadmm1}
\end{figure*}

The proposed distributed ADMM design iteratively updates auxiliary variables, BS precoding, RIS reflection coefficient vectors, and multipliers. However, it has high computational complexity since the conventional distributed ADMM may take hundreds or thousands of iterations to achieve convergence. System performance and convergence are additionally hampered by the requirement to manually choose crucial hyper-parameters, such as the power normalized factor $\left\{ {{\mu _b}|\forall b \in {\cal B}} \right\}$ and the penalty factor $\left\{ {{\rho _b}|\forall b \in {\cal B}} \right\}$. To overcome these shortcomings, we unfold the proposed distributed ADMM design into the D$^2$-ADMM to learn the hype-parameters $\left\{ {{\rho _b}|\forall b \in {\cal B}} \right\}$ automatically and bypass $\left\{ {{\mu _b}|\forall b \in {\cal B}} \right\}$. Besides, we create a neural block called $\theta$-Block to solve the complicated problem (\ref{eq20d}). The specific D$^2$-ADMM structure is illustrated in Fig. \ref{deepadmm1}.
\begin{figure*}[tbp]
\centering {
	\begin{tabular}{ccc}
		\includegraphics[width=0.98\textwidth]{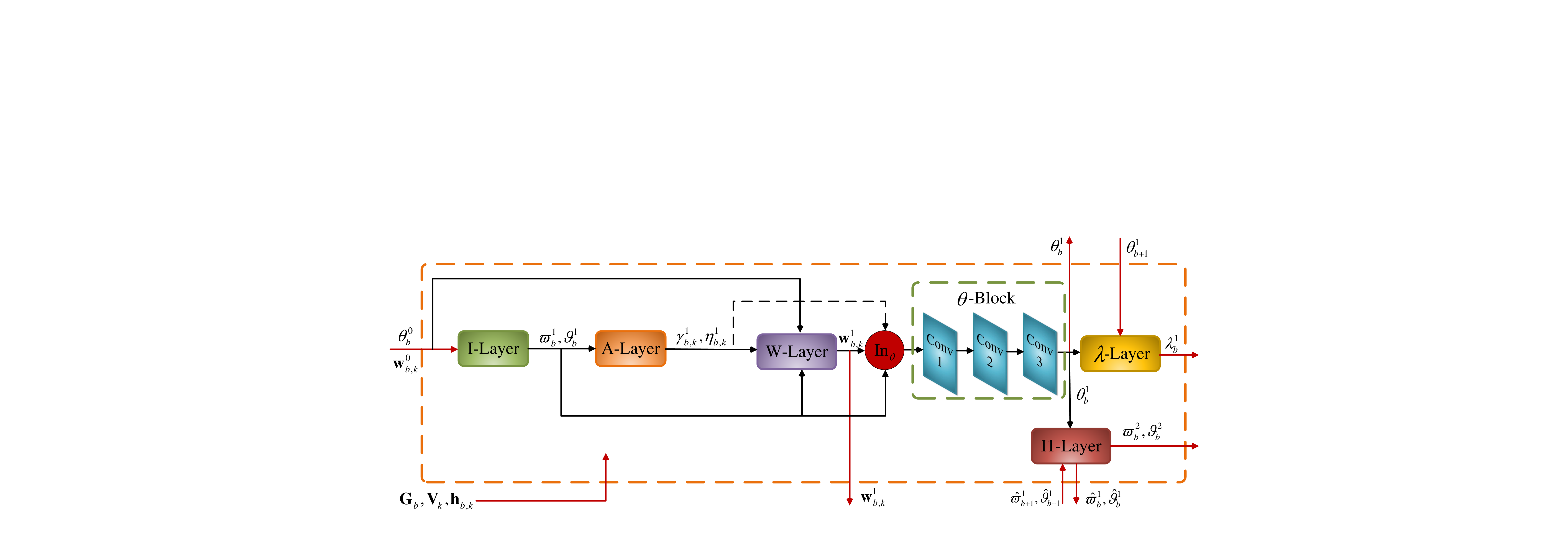}\\
	\end{tabular}
}
\caption{The structure of Ini-Block relying on algorithm unrolling.}
\vspace{-1\baselineskip}
\label{deepadmm2}
\end{figure*}

As shown in Fig. \ref{deepadmm1}, a total of $B$ D$^2$-ADMM are respectively implemented at $B$ BSs. A D$^2$-ADMM is composed of $L \ge B $ cascaded neural blocks. Each neural block is designed according to one iteration of the distributed ADMM design, which means that a neural block is equivalent to a single iteration in traditional iterative algorithms. The $(l-1)$-st neural block's output constitutes the input of the $l$-th neural block. The input of the first neural block is initialized, and the last neural block outputs the optimized BS precoding matrix and RIS reflection coefficient vectors.

More specifically, we have five different kinds of neural blocks, namely the initialization neural block (Ini-Block), the middle neural block 1 (Mid-Block1), the middle neural block 2 (Mid-Block2), the middle neural block 3 (Mid-Block3), and the output neural block (Out-Block). For the sake of illustration, we give the schematic diagram of the Ini-Block in Fig. \ref{deepadmm2}. The structures of other neural blocks are based on the Ini-Block by replacing or pruning certain parts. Ini-Block initializes the network, which includes a cross-term information initialization layer ($I$-Layer), an auxiliary variable update layer ($A$-Layer), a BS precoding update layer ($W$-Layer), an RIS update block ($\theta$-Block), a multiplier update layer ($\lambda$-Layer), and a cross-term information exchange layer 1 (I1-Layer). The first neural block of D$^2$-ADMM is an Ini-Block. The 2nd to $(B-1)$-st network blocks of D$^2$-ADMM are created as the Mid-Block1, which contains an $A$-Layer, a $W$-Layer, a $\theta$-Block, a $\lambda$-Layer, and a I1-Layer. The $B$-th neural block of D$^2$-ADMM is Mid-Block2, which has a similar structure as Mid-Block1, except that I1-Layer is replaced with a cross-term information exchange layer 2 (I2-Layer). Moreover, the $(B+1 \sim L-1)$-st neural blocks are Mid-Block3, which is constructed similarly as Mid-Block2 with the exception of using a cross-term information exchange layer 3 (I3-Layer). The last neural block of D$^2$-ADMM is referred to Out-Block, which consists of an $A$-Layer, a $W$-Layer, and a $\theta$-Block.

Next, we will discuss the structure and function of each layer and the $\theta$-Block.

\subsubsection{Auxiliary Variable Update Layer ($A$-Layer)} 
$A$-Layer updates two auxiliary variables, $\boldsymbol \gamma$ and $\boldsymbol \eta$, according to (\ref{eq21}) and (\ref{eq23}). To reflect the iteration order, we rewrite (\ref{eq21}) and (\ref{eq23}) as
\begin{align}   \label{eq31}
\gamma _{b,k}^l = \frac{{{{\left| {\varpi _{b,k,k}^{l} + \vartheta _{b,k,k}^{l}} \right|}^2}}}{{\sum\limits_{j = 1,j \ne k}^K {{{\left| {\varpi _{b,k,j}^{l} + \vartheta _{b,k,j}^{l}} \right|}^2} + {\delta ^2}} }},
\end{align}
\begin{align}   \label{eq32}
\eta _{b,k}^l = \frac{{{{\left( {\varpi _{b,k,k}^{l} + \vartheta _{b,k,k}^{l}} \right)}^ * }\sqrt {\left( {1 + \gamma _{b,k}^l} \right){\omega _k}} }}{{\sum\limits_{j = 1}^K {{{\left| {\varpi _{b,k,j}^{l} + \vartheta _{b,k,j}^{l}} \right|}^2} + {\delta ^2}} }},
\end{align}
respectively, where ${\varpi _{b,k,j}^{l}}$ and ${\vartheta _{b,k,j}^{l}}$ denote the cross-term information of the $l$-th neural block for the $b$-th BS; $\gamma _{b,k}^l$, and $\eta _{b,k}^l$ are the two auxiliary variables of the $l$-th neural block for the $b$-th BS.

\subsubsection{BS Precoding Update Layer ($W$-Layer)}
According to (\ref{eq26}), $W$-Layer updates the BS precoding matrix $\bf W$ as 
\begin{align}   \label{eq33}
{\bf{w}}_{b,k}^l = \frac{{\sqrt {\left( {1 + \gamma _{b,k}^l} \right){\omega _{b,k}}} {{\left( {\eta _k^l} \right)}^ * }{\bf{\hat h}}_{b,k}^{l - 1} - {\boldsymbol \Omega} _{b,k}^{l - 1}}}{{\sum\limits_{j = 1}^K {{{\left| {\eta _{b,j}^l} \right|}^2}} {\bf{\hat h}}_{b,j}^{l - 1}{{\left( {{\bf{\hat h}}_{b,j}^{l - 1}} \right)}^H}}},
\end{align}
where ${\boldsymbol \Omega} _{b,k}^{l - 1} = \sum\limits_{j = 1}^K {{{\left| {\eta _{b,j}^l} \right|}^2}{\bf{\hat h}}_{b,k}^{l - 1}} (\varpi _{b,j,k}^l + \vartheta _{b,j,k}^l - {({\bf{\hat h}}_{b,j}^{l - 1})^H}{\bf{w}}_{b,k}^{l - 1})$ and ${\bf{\hat h}}_{b,j}^{l - 1} $ $ = ( {{\bf{h}}_{b,j}^H + {{( {{\boldsymbol \theta} _b^{l - 1}} )}^H}{\bf{V}}_k^H{{\bf{G}}_b}} )^H$. 

In order to satisfy the power constraint, ${\bf{w}}_{b,k}^l$ can be rewrited as
\begin{align}   \label{eq34}
{\bf{w}}_{b,k}^{{l}} = \frac{{\sqrt {{P_{b,\max }}} {\bf{w}}_{b,k}^{l}}}{{\sqrt {\sum\limits_{k = 1}^K  {\left\| {{{\bf{w}}_{b,k}^l}} \right\|}^2} }}.
\end{align}

\subsubsection{RIS Update Block ($\theta$-Block)} 
As previously mentioned, (\ref{eq27}) is a non-convex function that is challenging to solve by conventional methods. Therefore, we introduce the $\theta$-Block, which aims to exploit the inference ability of DL to solve this problem. $\theta$-Block is composed of multiple convolutional layers. Specifically, we first rewrite (\ref{eq27}) as
\begin{align}   \label{eq35}
\angle \left( {{{\boldsymbol{\theta }}_b}} \right) = {f_\theta }\left( {{\bf{S}},{\bf{Z}}} \right),
\end{align}
where ${f_\theta }$ denotes a non-linear function that applied as the solver of problem (\ref{eq27}).

We then use multiple convolutional layers to approximate this complicated non-linear function ${f_\theta }$. Since the neural network is more amenable with real-valued data, we first convert $\bf{S}$ and $\bf{Z}$ into real-valued sequences ${\rm{I}}{{\rm{n}}_\theta }$ as the input of the $\theta$-Block, expressed as follows.
\begin{align}   \label{eq36}
{\rm{I}}{{\rm{n}}_\theta } = \left[ {{\mathop{\rm Re}\nolimits} \left\{ {\bf{S}} \right\},{\mathop{\rm Im}\nolimits} \left\{ {\bf{S}} \right\},{\mathop{\rm Re}\nolimits} \left\{ {\bf{Z}} \right\},{\mathop{\rm Im}\nolimits} \left\{ {\bf{Z}} \right\}} \right].
\end{align}

Therefore, the working principle of $\theta$-Block can be expressed as
\begin{align}   \label{eq37}
\angle \left( {{{\boldsymbol {\theta }}_b}} \right) = {f_{C,U}}\left( { \cdots {f_{C,u}}\left( { \cdots {f_{C,1}}\left( {{\rm{I}}{{\rm{n}}_\theta }|{\upsilon _1}} \right)|{\upsilon _u}} \right)|{\upsilon _U}} \right),
\end{align}
where ${{f_{C,u}}}$ is the $u$-th convolutional layer; $U$ denotes the number of convolutional layers; ${{\upsilon _u}}$ is the parameter set of the $u$-th convolutional layer. In this paper, we empirically choose $U=3$ which is sufficient for our problem.

Note that in the proposed architecture, the parameters of each convolutional layer can be automatically learned through end-to-end training.

\subsubsection{Multiplier Update Layer ($\lambda$-Layer)}
The multipliers are updated through this layer using the following strategy
\begin{align}   \label{eq38}
{\boldsymbol{\lambda }}_b^l = {\boldsymbol{\lambda }}_b^{l - 1} + {\rho _b}\left( {{\boldsymbol{\theta }}_b^l - {\boldsymbol{\theta }}_{_{\bar b}}^l} \right),
\end{align}
where ${\rho _b}$ is a learnable parameter; ${\boldsymbol{\theta }}_{_{\bar b}}^l$ is the RIS reflection coefficient vector exchanged from the ${\bar b}$-th BS.
\subsubsection{Cross-Term Information Initialization Layer (I-Layer)}
Again, in the cooperative design of distributed RIS-assisted cell-free systems, CSI sharing is necessary among BSs. However, considering the security and the excessive overhead associated with direct CSI exchange, we define $\left\{ {{\varpi _{b,k,j}},{\vartheta _{b,k,j}}|\forall b \in {\cal B};\forall k,j \in \cal K} \right\}$ as two types of necessary cross-information in $A$-layer, $W$-Layer, and $\theta$-Block. 

I-Layer initializes the local cross-term information, which is expressed as
\begin{subequations} \label{eq39}
\begin{align}   
	&\left\{ \begin{array}{l}
		\hat \varpi _{b,k,j}^0 = 0,\\
		\hat \vartheta _{b,k,j}^0 = 0,
	\end{array} \right.\label{eq39a}\\
	&\left\{ \begin{array}{l}
		\varpi _{b,k,j}^1 = {\bf{h}}_{b,k}^H{\bf{w}}_{b,k}^0,\\
		\vartheta _{b,k,j}^1 = {\left( {\boldsymbol{\theta} _b^0} \right)^H}{\bf{V}}_k^H{{\bf{G}}_b}{\bf{w}}_{b,k}^0,
	\end{array} \right.\label{eq39b}			
\end{align}
\end{subequations}
where $\hat \varpi _{b,k,j}^0$ and $\hat \vartheta _{b,k,j}^0$ are two initialized cross-term information, which will be sent to the adjacent BSs; $\varpi _{b,k,j}^1$ and $\vartheta _{b,k,j}^1$ denote two cross-term information, which will be used for updating the next neural block; ${\bf{w}}_{b,k}^0$ and ${\boldsymbol{\theta} _b^0} $ are initialized randomly.

\subsubsection{Cross-Term Information Layer 1 (I1-Layer)} 
I1-Layer includes two processes. The process 1 is to send the updated cross-term information to the adjacent BSs, expressed as (\ref{eq40a}), where $\hat \varpi _{b,k,j}^l$ and $\hat \vartheta _{b,k,j}^l$ are two cross-term information that needs to be shared with the adjacent BSs. Moreover, $\hat \varpi _{\bar b,k,j}^{l - 1}$ and $\hat \vartheta _{\bar b,k,j}^{l - 1}$ represent two cross-term information symbols that are received from the adjacent BSs. ${\boldsymbol{\theta} _b^l}$ and ${\bf{w}}_{b,k}^l$ denote the $l$-th update of the RIS reflection coefficient vector and the BS precoding vector, respectively. In the process 2, the cross-term information required for updating the next neural block will be determined based on the received cross-term information from the adjacent BSs, as demonstrated in (\ref{eq40b}), where $\varpi _{b,k,j}^{l + 1}$ and $\vartheta _{b,k,j}^{l + 1}$ denote the two cross-term information symbols that are required for updating the $(l+1)$-st neural block. I1-Layer is configured for the $\left( {1 \sim B{ - 1}} \right)$-st neural blocks.
\begin{subequations} \label{eq40}
\begin{align}   
	&\left\{ \begin{array}{l}
		\hat \varpi _{b,k,j}^l = \hat \varpi _{\bar b,k,j}^{l - 1} + {\bf{h}}_{b,k}^H{\bf{w}}_{b,k}^l,\\
		\hat \vartheta _{b,k,j}^l = \hat \vartheta _{\bar b,k,j}^{l - 1} + {\left( {\boldsymbol{\theta} _b^l} \right)^H}{\bf{V}}_k^H{{\bf{G}}_b}{\bf{w}}_{b,k}^l,
	\end{array} \right.\label{eq40a}\\
	&\left\{ \begin{array}{l}
		\varpi _{b,k,j}^{l + 1} = \hat \varpi _{\bar b,k,j}^l + {\bf{h}}_{b,k}^H{\bf{w}}_{b,k}^l,\\
		\vartheta _{b,k,j}^{l + 1} = \hat \vartheta _{\bar b,k,j}^l + {\left( {\boldsymbol{\theta} _b^l} \right)^H}{\bf{V}}_k^H{{\bf{G}}_b}{\bf{w}}_{b,k}^l.
	\end{array} \right.\label{eq40b}			
\end{align}
\end{subequations}

\subsubsection{Cross-Term Information Layer 2 (I2-Layer)}
I2-Layer has the similar process 1 but distinct process 2 as I1-Layer. Specifically, the updates of the $b$-th BS in the first neural block is included in the cross-term information needed for updating the $(B+1)$-st neural block. Thus, we have to eliminate the obsolete updates from the cross-term information and add the $B$-th update to guarantee that only the new update is included. The specific process 2 is expressed as follows
\begin{align}   \label{eq41}
\left\{ \begin{array}{l}
	\varpi _{b,k,j}^{B + 1} = \hat \varpi _{\bar b,k,j}^B - {\bf{h}}_{b,k}^H{\bf{w}}_{b,j}^{1} + {\bf{h}}_{b,k}^H{\bf{w}}_{b,j}^B,\\
	\vartheta _{b,k,j}^{B + 1} = \hat \vartheta _{\bar b,k,j}^B - {\left( {\boldsymbol{\theta} _b^{1}} \right)^H}{\bf{V}}_k^H{{\bf{G}}_b}{\bf{w}}_{b,j}^{1} + {\left( {\boldsymbol{\theta} _b^B} \right)^H}{\bf{V}}_k^H{{\bf{G}}_b}{\bf{w}}_{b,j}^B.
\end{array} \right.
\end{align}

Therefore, I2-Layer is only exploited in the $B$-th neural block. 

\subsubsection{Cross-Term Information Layer 3 (I3-Layer)}
When $l \ge B + 1$, both the cross-term information to be sent to the adjacent BSs and the cross-term information used for updating the next neural block need to eliminate obsolete updates. Therefore, the two processes of I3-Layer can be described as
\begin{subequations} \label{eq42}
\begin{align}   
	&\left\{ \begin{array}{l}
		\hat \varpi _{b,k,j}^l = \hat \varpi _{\bar b,k,j}^{l - 1} - {\bf{h}}_{b,k}^H{\bf{w}}_{b,j}^{l - B + 1} + {\bf{h}}_{b,k}^H{\bf{w}}_{b,j}^l,\\
		\hat \vartheta _{b,k,j}^l = \hat \vartheta _{\bar b,k,j}^{l - 1} - {\left( {\boldsymbol{\theta} _b^{l - B + 1}} \right)^H}{\bf{V}}_k^H{{\bf{G}}_b}{\bf{w}}_{b,j}^{l - B + 1} + {\left( {\boldsymbol{\theta} _b^l} \right)^H}{\bf{V}}_k^H{{\bf{G}}_b}{\bf{w}}_{b,j}^l,
	\end{array} \right.\label{eq42a}\\
	&\left\{ \begin{array}{l}
		\varpi _{b,k,j}^{l + 1} = \hat \varpi _{\bar b,k,j}^l - {\bf{h}}_{b,k}^H{\bf{w}}_{b,j}^{l - B + 2} + {\bf{h}}_{b,k}^H{\bf{w}}_{b,j}^l,\\
		\vartheta _{b,k,j}^{l + 1} = \hat \vartheta _{\bar b,k,j}^l - {\left( {\boldsymbol{\theta} _b^{l - B + 2}} \right)^H}{\bf{V}}_k^H{{\bf{G}}_b}{\bf{w}}_{b,j}^{l - B + 2} + {\left( {\boldsymbol{\theta} _b^l} \right)^H}{\bf{V}}_k^H{{\bf{G}}_b}{\bf{w}}_{b,j}^l.
	\end{array} \right.\label{eq42b}			
\end{align}
\end{subequations}

We deploy I3-Layer in the $\left( {B+1 \sim L-1} \right)$-st neural blocks.

\subsection{Information Exchange Strategy}
\begin{figure}[tbp]
\centering {
	\begin{tabular}{ccc}
		\includegraphics[width=0.5\textwidth]{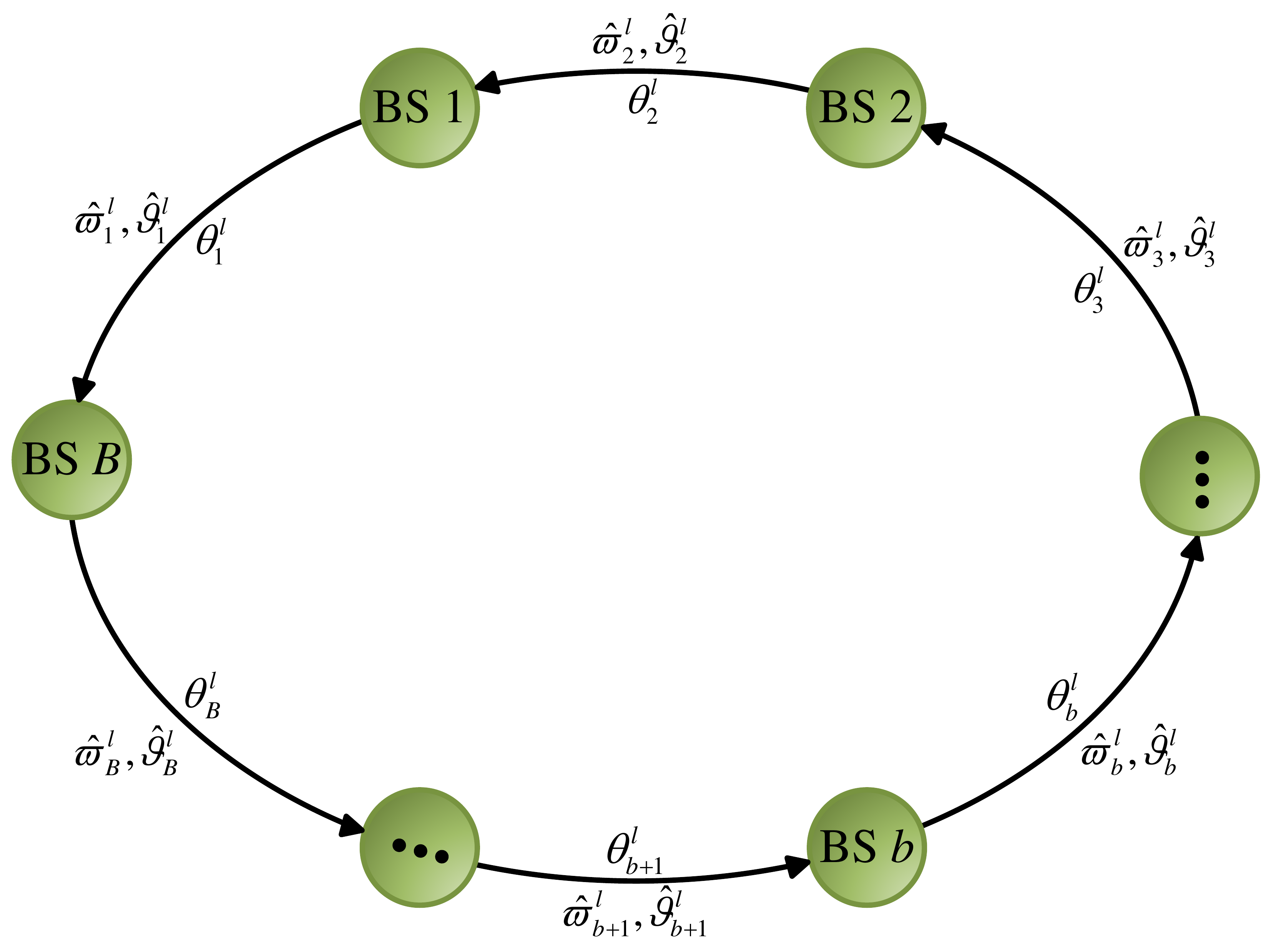}\\
	\end{tabular}
}
\caption{The proposed information exchange strategy.}
\vspace{-1\baselineskip}
\label{inforex}
\end{figure}
Next, we elaborate on the proposed information exchange strategy. To safeguard the information privacy of different BSs and reduce the proposed system's information exchange overhead, we define two types of cross-term information used for the update at each BS. The updating of each neural block needs to guarantee the integrality and timeliness of the cross-term information, as demonstrated by the updating process of the I1-layer, the I2-layer, and the I3-layer. In most existing distributed information exchange strategies, each BS often receives information shared by multiple BSs \cite{RIScellfree3,RIScellfree4}. This exchange strategy will reduce the integrality and timeliness of the cross-term information defined in our paper, affecting the convergence and performance of the system. Therefore, we propose an effective monodirectional information exchange strategy, assuming all BSs have a monodirectional topology, as illustrated in Fig. \ref{inforex}.

Each BS performs a monodirectional information exchange with two adjacent BSs through a dedicated link. For instance, the $b$-th BS receives cross-term information from the $(b+1)$-st BS and sends its cross-term information to the $(b-1)$-st BS. Such a strategy requires at least $B$ exchanges to ensure the integrality of the cross-term information. As the iteration proceeds, the timeliness of the cross-term information is guaranteed by replacing the obsolete information with the latest information. The specific cross-term information processing are completed at the I1-Layer, I2-Layer, and I3-Layer.

In addition to exchanging cross-term information, we also need to exchange the RIS reflection coefficient vectors updated by each neural block among various BSs to update the multiplier $\lambda$. Therefore, the $b$-th BS needs to send $\{ {\hat {\varpi} _{b,k,j}^l|\forall k,j \in \cal K} \}$ ($K^2$ dimension), $\{ {\hat{\vartheta}  _{b,k,j}^l|\forall k,j \in \cal K} \}$ ($K^2$ dimension), and ${\boldsymbol \theta} _b^l$ ($RN$ dimension) in the $l$-th neural block. As a consequence, the total dimension of exchanged data in the practical RIS-assisted  cell-free system is $B(L-1)(2K^2+RN)$, which is significantly reduced compared with that exchanging CSI directly.

\subsection{Training of D$^2$-ADMM}
In this section, we give the specific training and practical application methods of the proposed D$^2$-ADMM. The input to D$^2$-ADMM at the $b$-th BS is its local CSI, the initialized ${\bf{w}}_{b,k}^0$ and ${\boldsymbol{\theta} _b^0} $, while the output is the optimized ${\bf{w}}_{b,k}^L$ and ${\boldsymbol{\theta} _b^L} $. Then the parameters of $\theta$-Layer and $\rho _b$ in D$^2$-ADMM are updated through an end-to-end training. The loss function for training is set as 
\begin{align}   \label{eq43}
{f_{{\rm{Loss}}}} = \frac{1}{Q}\sum\limits_{q = 1}^Q {\sum\limits_{b = 1}^B {\underbrace {{{\left\| {{{\boldsymbol{\theta }}_{q,b}} - {{\boldsymbol{\theta }}_{q,\bar b}}} \right\|}^2}}_{{\rm{Consensus}}\;{\rm{error}}}}  - {\rm{WS}}{{\rm{R}}_q}} ,
\end{align}
where $Q$ is the sample number of one training batch.

By minimizing the loss function ${f_{{\rm{Loss}}}}$, the consensus error is minimized while maximizing WSR. It is worth noting that the training process is completed on a single CPU. After completing the training, we deploy $B$ D$^2$-ADMMs to the corresponding BSs for practical distributed implementation.

%

\section{Numerical Results}
This section provides simulation results to demonstrate the effectiveness of our proposed D$^2$-ADMM framework for the RIS-assisted cell-free system.

\subsection{Simulation Setup}
\begin{figure}[tbp]
\centering {
	\begin{tabular}{ccc}
		\includegraphics[width=0.5\textwidth]{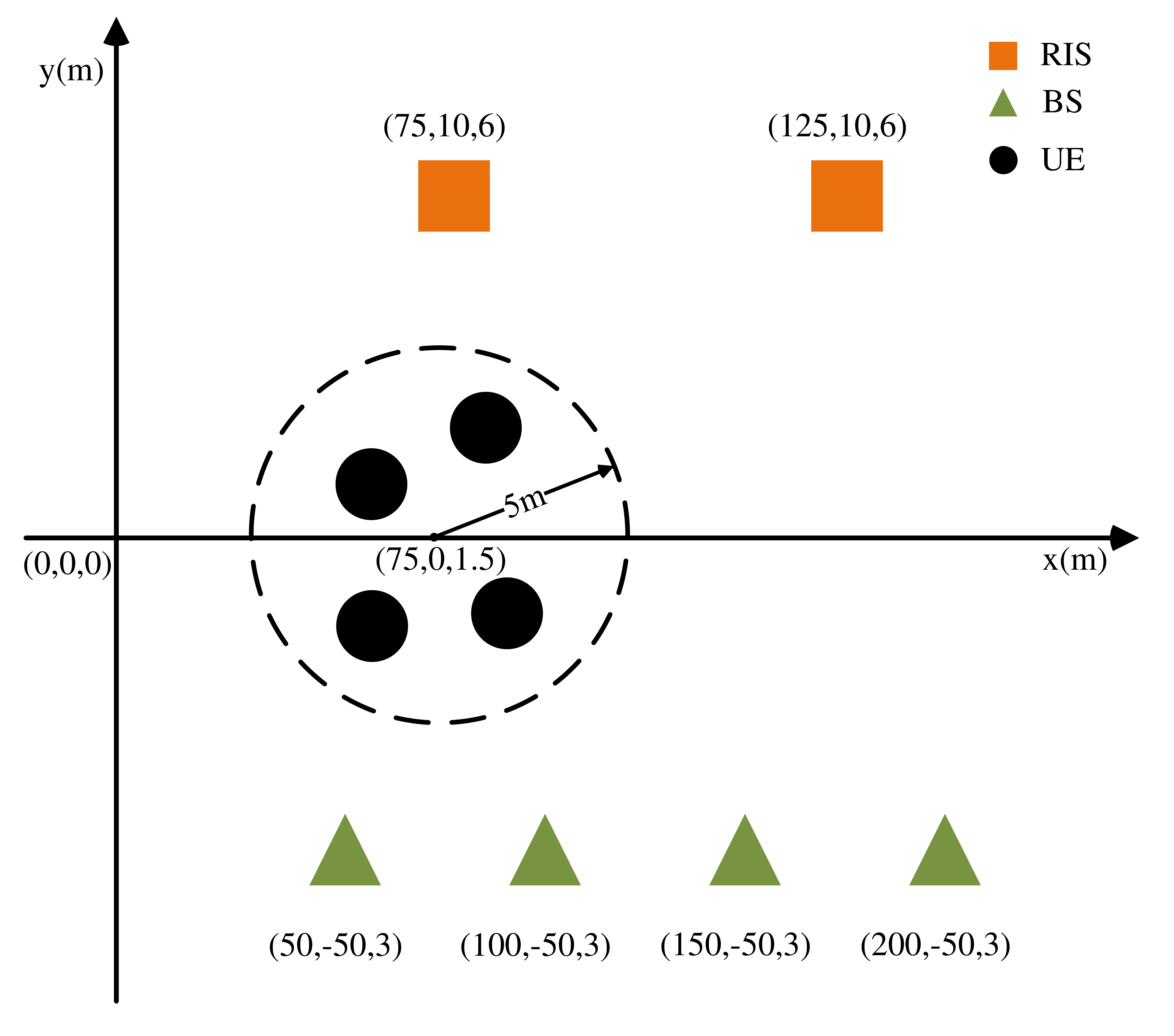}\\
	\end{tabular}
}
\caption{The 3D scenario of the RIS-assisted cell-free system.}
\vspace{-1\baselineskip}
\label{3dset}
\end{figure}
We consider a typical RIS-assisted cell-free system 3D scenario shown in Fig. \ref{3dset}. In this scenario, the $b$-th BS is deployed at $(200 \times \frac{b}{B}, -50, 3)$ m. Without loss of generality, we consider $R=2$ RISs, which are deployed at $(75, 10, 6) $m and $(125,10,6)$ m. $K$ UEs served by $B$ BSs are randomly distributed in a circular area with a center at $(75, 0, 1.5)$ m, a radius of $5$m, and a height of $1.5$ m. The number of antennas at each BS is set to $N_t=2$. Given the location information of each device, the corresponding channel can be determined by (\ref{eq1}). In this setup, we assume that the multi-path number of each channel is $3$ ($1$ LoS, $2$ NLoS), and their AOAs and AODs are chosen randomly in the range $[ { - \frac{\pi }{2},\frac{\pi }{2}} ]$. Likewise, all BSs have the same maximum transmit power, i.e., ${P_{b,\max }} = P$. The received noise power is set to ${\delta ^2} = -80$ dBm.

To better demonstrate the performance of the proposed D$^2$-ADMM, we consider several representative benchmarks, as listed below.
\begin{itemize}
\item \emph{Centralized}: Assuming that all BSs send their local CSI to the central CPU for the centralized design of the BS precoding matrix and the RIS reflection coefficient vectors \cite{RIScellfree4}. 
\item \emph{MRT Random $\theta$}: A distributed design method, where the RIS reflection coefficient vector is randomly configured, and the precoding of BS is designed as the conjugate of local CSI \cite{cellfree1}.
\item \emph{MRT Comb MaxAO}: A distributed algorithm that maximizes the channel gain of cascaded channels for configuring the RIS, and the design of BS precoding is the same as MRT Random $\theta$.
\item \emph{Local ZF Comb MaxAO}: This distributed algorithm has the same design of RIS as MRT Comb MaxAO and exploits the local ZF algorithm proposed in \cite{LocalZF} for optimizing the BS precoding matrix.
\end{itemize}

\subsection{Training Performance of D$^2$-ADMM}
\begin{figure*}[tbp]
\centering
\vspace{-1\baselineskip}
\subfigure[]{
	\label{R2}
	\begin{minipage}[t]{0.31\linewidth}
		\includegraphics[width=1\textwidth]{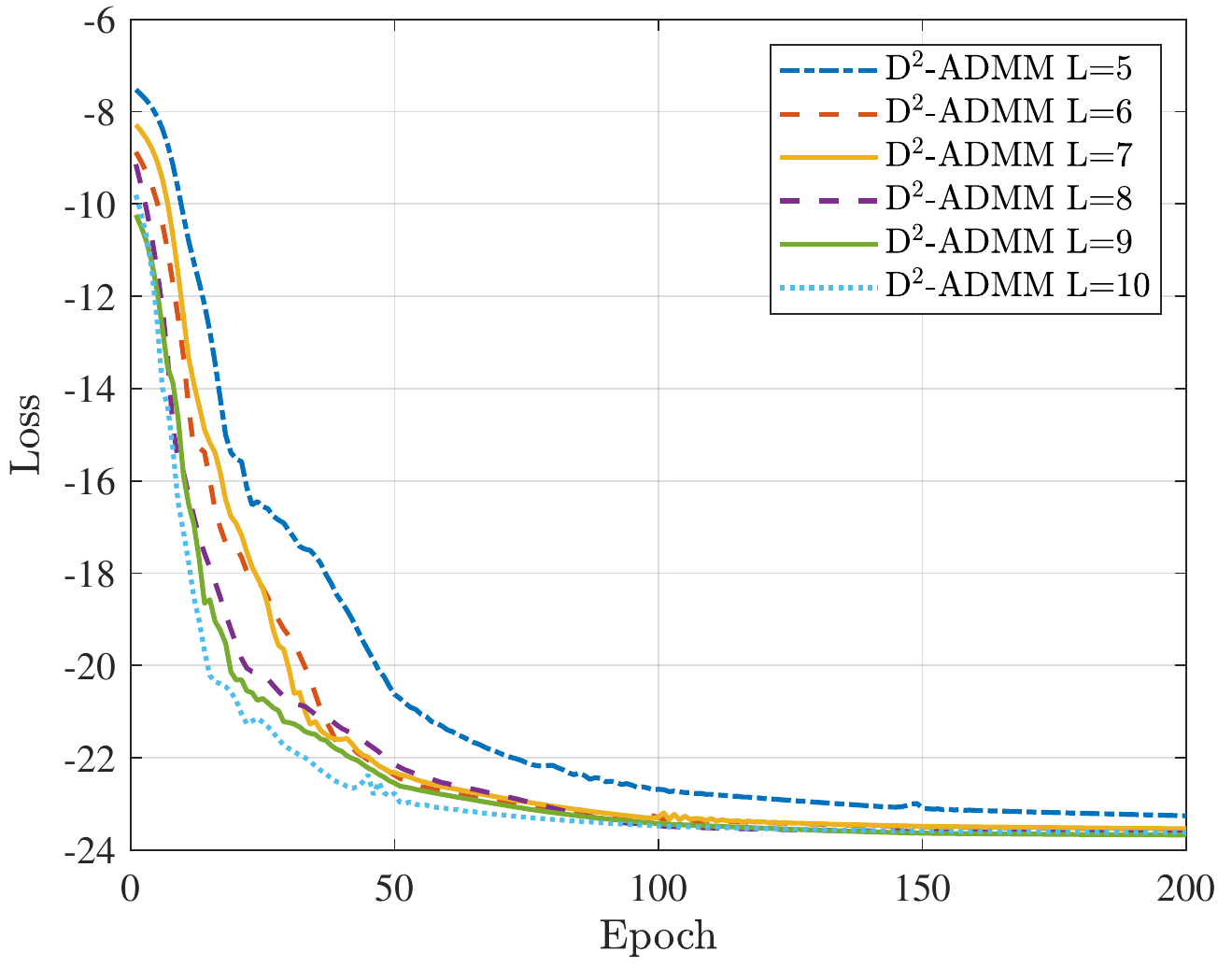}\notag
	\end{minipage}
}
\subfigure[]{
	\label{R3}
	\begin{minipage}[t]{0.31\linewidth}
		\includegraphics[width=1\textwidth]{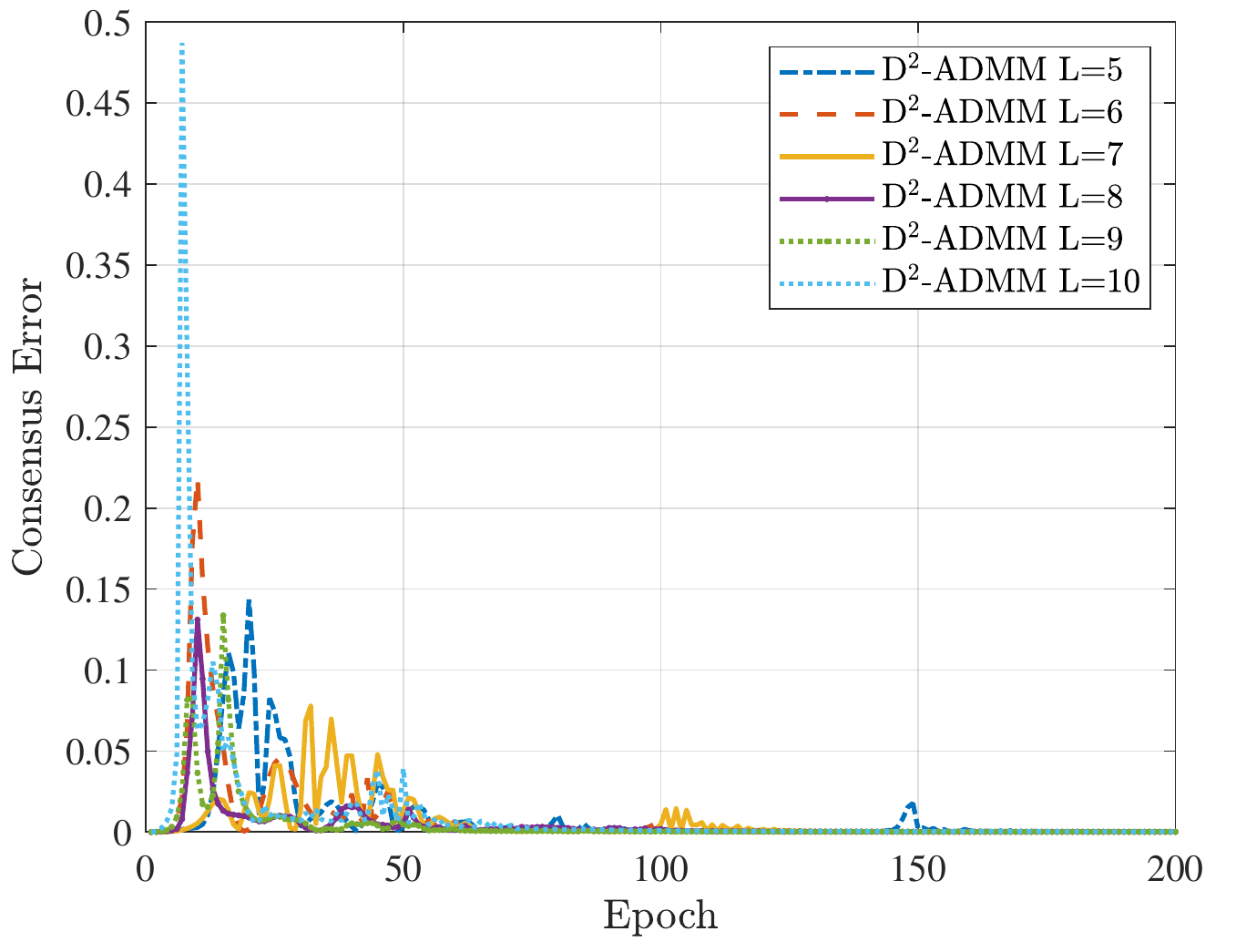}\notag
	\end{minipage}
}
\subfigure[]{
	\label{R1}
	\begin{minipage}[t]{0.31\linewidth}
		\includegraphics[width=1\textwidth]{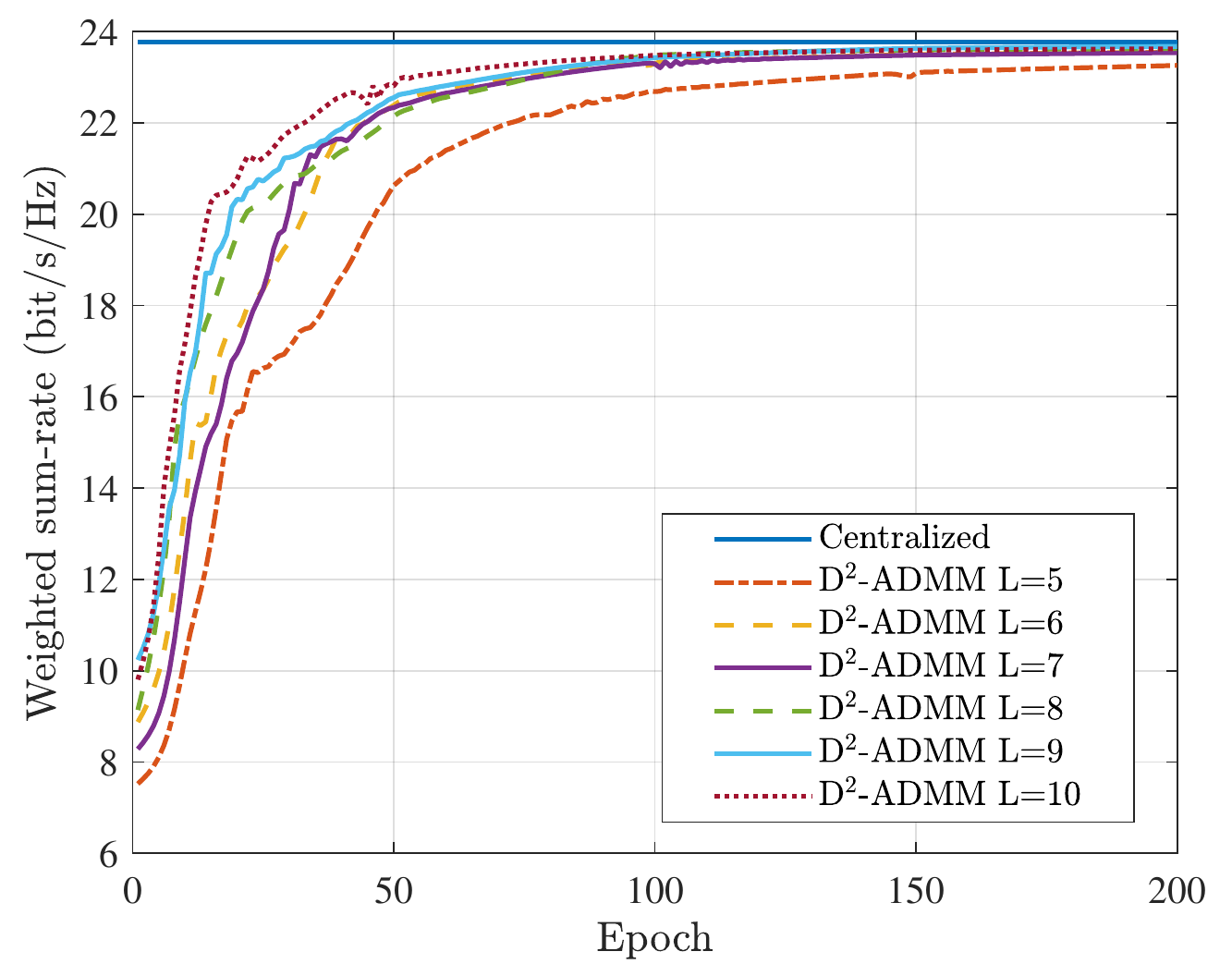}
	\end{minipage}
}
\caption{(a) The training loss of D$^2$-ADMM; (b) The consensus error of D$^2$-ADMM in the training process; (c) The WSR of D$^2$-ADMM in the training process;}
\vspace{-1\baselineskip}
\end{figure*}

In order to show the convergence of D$^2$-ADMM, we first conduct experiments to evaluate various indicators in the training process of D$^2$-ADMM, as shown in Figs. \ref{R2}-\ref{R1}, where we set $B=4$, $N=50$, $K=4$, $P=30$ dBm. 

Specifically, Fig. \ref{R2} illustrates the training loss of the D$^2$-ADMM under the different number of neural blocks. It can be seen that the D$^2$-ADMM training loss can converge as the training proceeds. In addition, the final convergent training loss gap for different $L$ is negligible when $L \ge 6$.
Furthermore, Fig. \ref{R3} shows the fluctuation of the consensus error of D$^2$-ADMM against different $L$. From Fig. \ref{R3}, we can observe that the consensus error of D$^2$-ADMM can converge to a minimal value as the training proceeds, and varied $L$ does not severely impact the convergence result of the consensus error.
The WSR in the training phase against the number of neural blocks is plotted in Fig. \ref{R1}, demonstrating that D$^2$-ADMM can gradually converge to performance comparable to the centralized algorithm as the training progresses. Moreover, D$^2$-ADMM converges more quickly as the number of neural blocks increases. Again, the final convergence performance reaches saturation when $L \ge 6$ in the simulation setups considered.

By comparing Fig. \ref{R2}-\ref{R1}, it can be concluded that the performance of D$^2$-ADMM can converge nearly to that of the centralized algorithm and gradually saturate as the number of neural blocks $L$ grows. Considering the tradeoff between the number of neural blocks and system performance, we provide a empirical selection criterion for the number of neural blocks as $L=B+2$.

\subsection{Performance of D$^2$-ADMM under Various Setups}
\begin{figure}[tbp]
	\centering {
		\begin{tabular}{ccc}
			\includegraphics[width=0.5\textwidth]{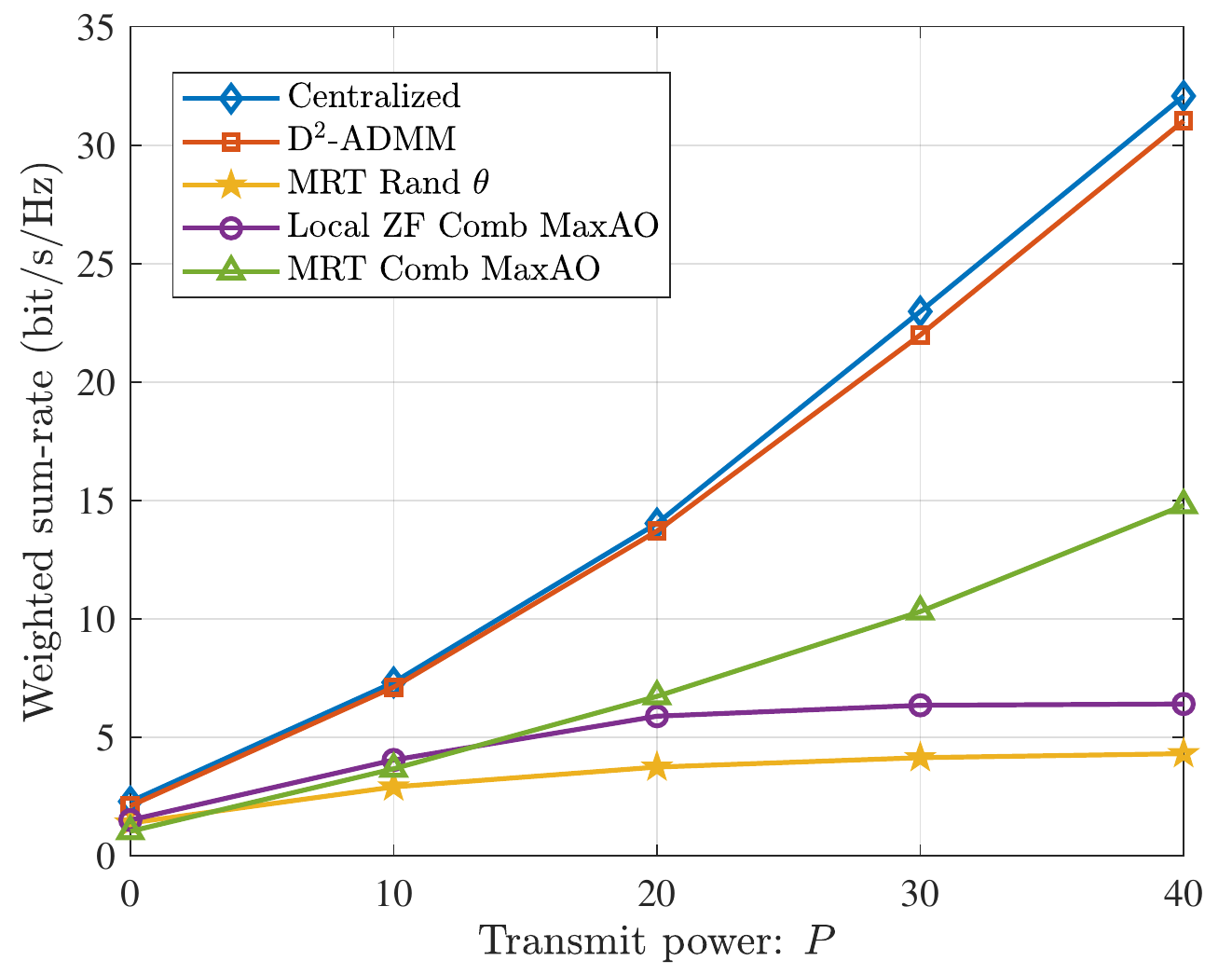}\\
		\end{tabular}
	}
	\caption{The WSR comparison against transmit power $P$, where $B=4$, $N_t=2$, $N=50$, $K=4$.}
	\vspace{-1\baselineskip}
	\label{R4}
\end{figure}
This section presents the performance comparison of D$^2$-ADMM and benchmark algorithms under various setups.

In Fig. \ref{R4}, we compare the WSR against the transmit power $P$ of different algorithms when $B=4$, $N=50$, $K=4$. According to the conclusions given in Section V-A, we choose $L = 6$ to balance the computational complexity and system performance. As shown in Fig. \ref{R4}, the WSR of all algorithms increases as $P$ increases. The centralized algorithm performs the best because it perfectly utilizes the CSI of all BSs. D$^2$-ADMM is demonstrated to have comparable performance, e.g., $95.6\%$ when $P=30$ dBm, to the centralized algorithm. The MRT Rand $\theta$ algorithm performs the worst because the unoptimized RIS reflection coefficient does not attain any benefits. Since Local ZF Comb MaxAO and MRT Comb MaxAO algorithms are non-distributed algorithms without incorporating all BSs for system design, they suffer from severe performance penalty compared with the proposed D$^2$-ADMM, e.g., the WSR by applying the D$^2$-ADMM attains about $213\%$ WSR improvement compared with the Local ZF Comb MaxAO when $P=30$ dBm.
\begin{figure}[tbp]
\centering {
	\begin{tabular}{ccc}
		\includegraphics[width=0.5\textwidth]{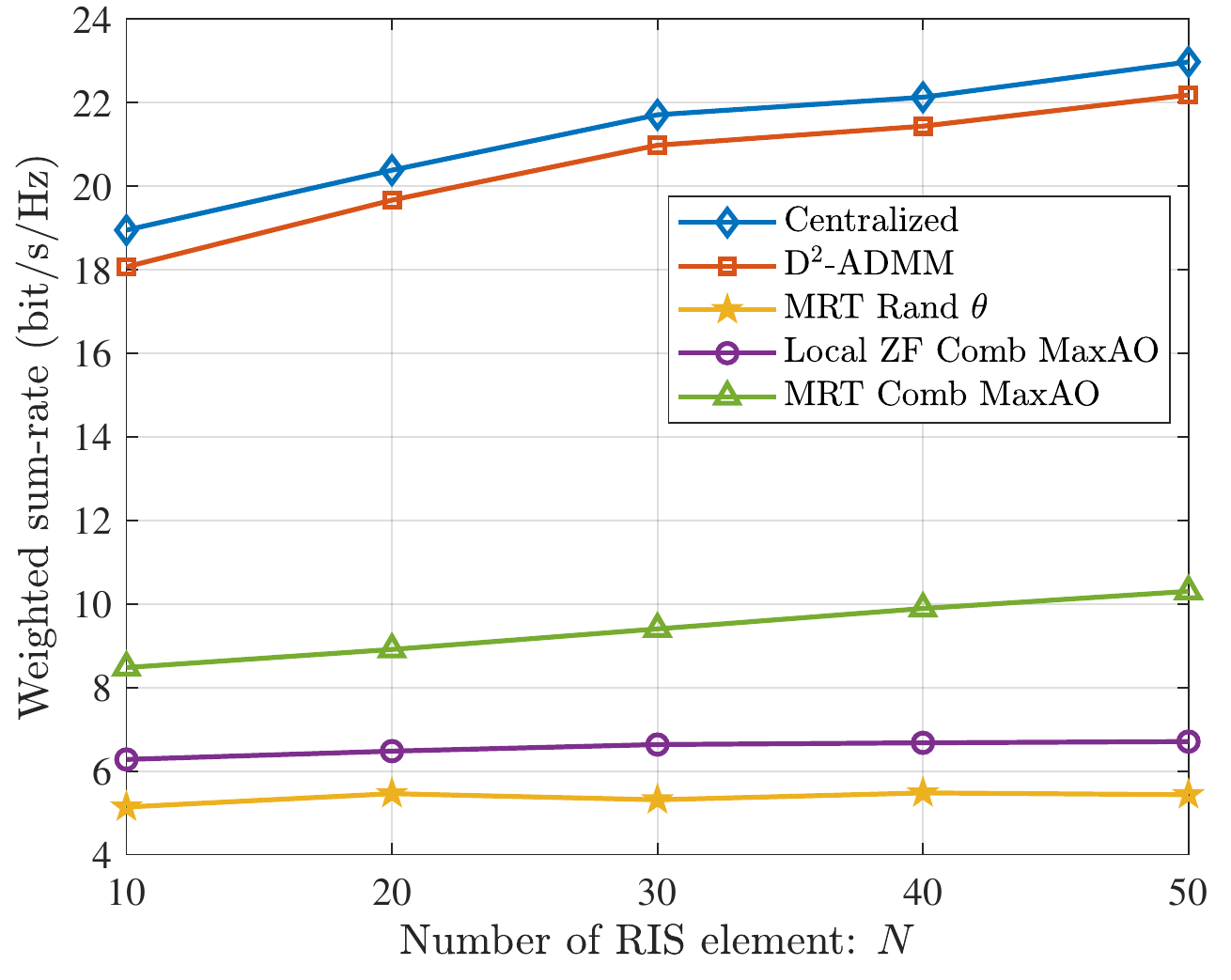}\\
	\end{tabular}
}
\caption{The WSR comparison against the number of RIS elements $N$, where $B=4$, $N_t=2$, $K=4$, $P=30$ dBm.}
\vspace{-1\baselineskip}
\label{R5}
\end{figure}

Fig. \ref{R5} shows the performance comparison between D$^2$-ADMM and benchmarks for different number of RIS elements $N$, where $B=4$, $K=4$, $P=30$ dBm. Observe from Fig. \ref{R5} that the centralized algorithm, the D$^2$-ADMM, and the local ZF Comb MaxAO algorithms improve as $N$ increases. However, MRT Comb MaxAO and MRT rand $\theta$ algorithms hardly benefit from increasing the number of  RIS elements. Besides, D$^2$-ADMM outperforms the other three distributed design algorithms, e.g., $223\%$ compared with the Local ZF Comb MaxAO when $N=30$, and can attain comparable performance, e.g., $96.6\%$ when $N=30$, to the centralized method. 

\begin{figure}[tbp]
\centering {
	\begin{tabular}{ccc}
		\includegraphics[width=0.5\textwidth]{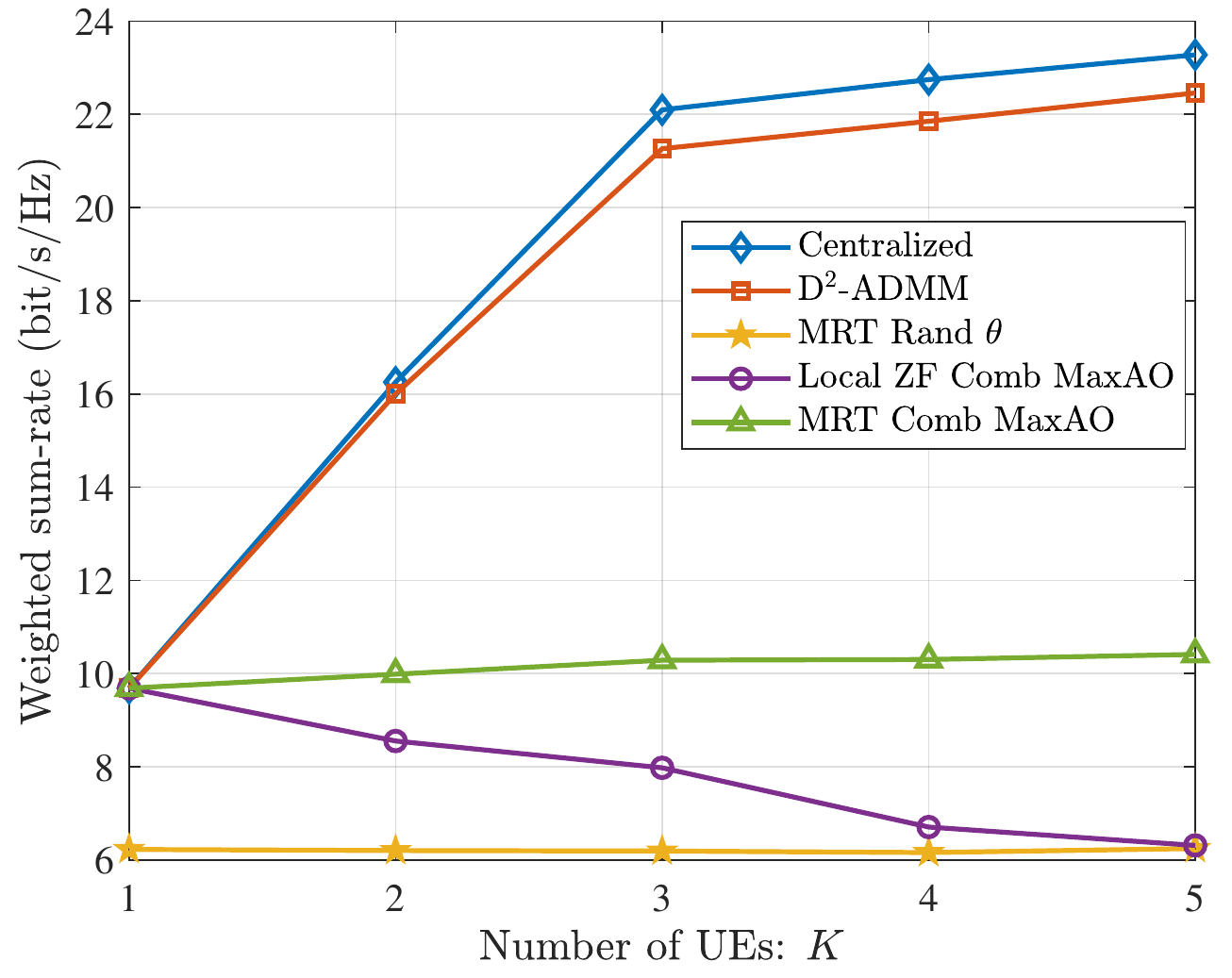}\\
	\end{tabular}
}
\caption{The WSR comparison against the number of UE $K$, where $B=4$, $N_t=2$, $N=50$, $P=30$ dBm.}
\vspace{-1\baselineskip}
\label{R6}
\end{figure}

Next, we show the WSR of various algorithms versus the number of UEs in Fig. \ref{R6}, where $B=4$, $N=50$, $P=30$  dBm. The centralized algorithm, the D$^2$-ADMM, and the Local ZF Comb MaxAO algorithm increase with $K$ thanks to the spatial multiplexing gain brought by the increased number of UEs. Again, the D$^2$-ADMM can perform as well as the centralized algorithm, e.g., about $96.5\%$ when $K=5$, and better than the Local ZF Comb MaxAO algorithm, e.g., about $216\%$ when $K=5$. When only a single UE is served, the performance of the other four algorithms is the same except for the MRT rand $\theta$ algorithm since the inter-user interference disappears in this situation. However, as a larger number of UEs access into the network, the MRT Comb MaxAO algorithm's performance declines, due to the fact that the distributed algorithm fails to suppress the inter-user interference.

\begin{figure}[tbp]
	\centering {
		\begin{tabular}{ccc}
			\includegraphics[width=0.5\textwidth]{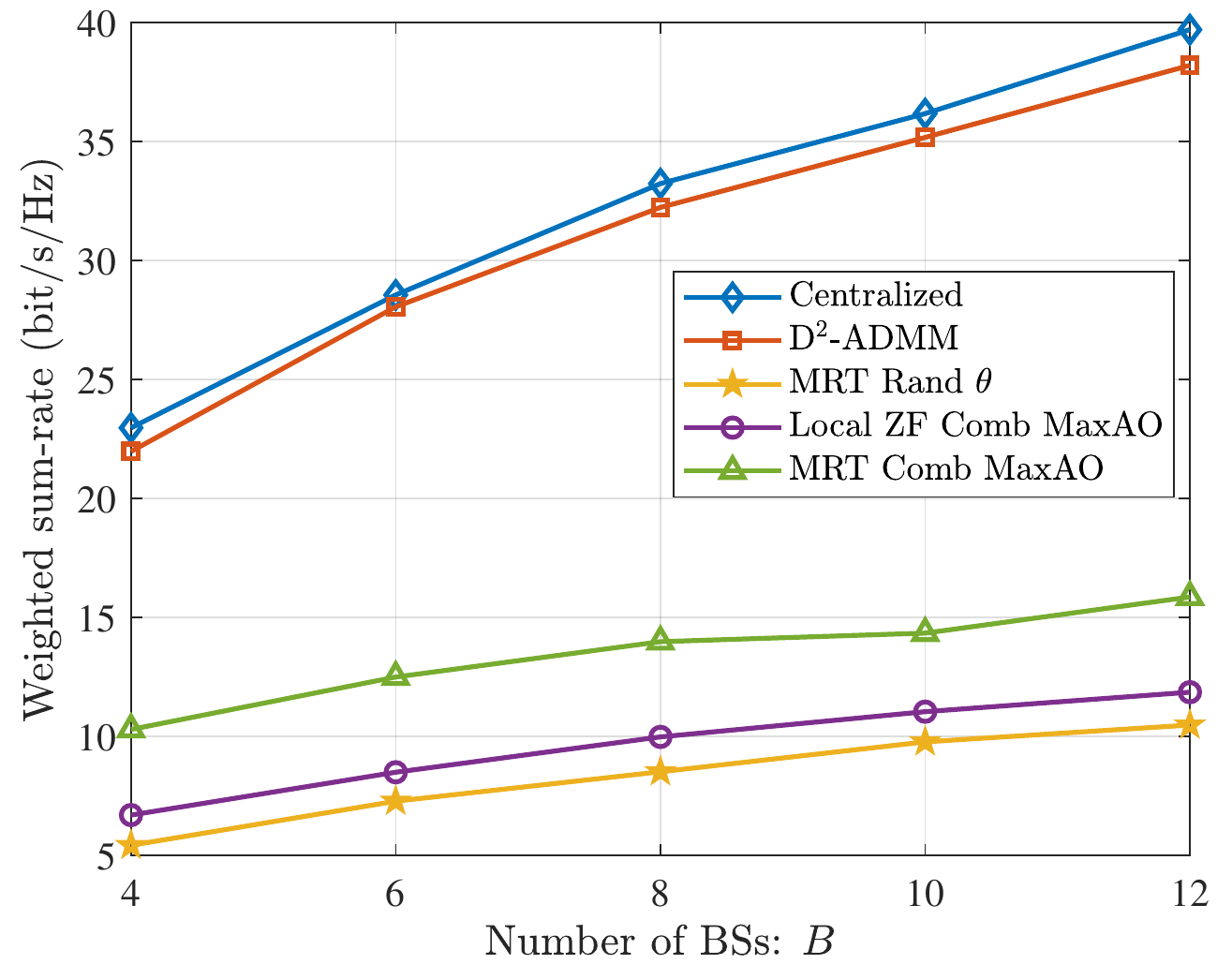}\\
		\end{tabular}
	}
	\caption{The WSR comparison against the number of BS $B$, where $N_t=2$, $N=50$, $K=4$, $P=30$ dBm.}
	\vspace{-1\baselineskip}
	\label{R7}
\end{figure}

Finally, we evaluate the D$^2$-ADMM algorithm's performance against other benchmarks by considering various numbers of BSs $B$ in Fig. \ref{R7}, where $N=50$, $K=4$, $P=30$ dBm. Again, the D$^2$-ADMM can achieve comparable performance, e.g., about $96.2\%$ when $B=5$, to the centralized algorithm with different $B$. The performance of D$^2$-ADMM also increases as $B$ increases since that more BSs can provide more power for UEs.

\section{Conclusion}
In this paper, we considered a RIS-assisted cell-free system that can boost communication capacity and overcome the drawbacks of conventional cellular networks. To jointly design the downlink precoding of BSs and the reflection phase shifts of RISs,  we proposed a distributed cooperative design based on ADMM, which can fully utilize the parallel computing resources. Subsequently, we developed a neural network framework, D$^2$-ADMM, by unrolling each iteration of the proposed distributed cooperative design, to automatically learn hyper-parameters and non-convex RIS solvers through end-to-end training. Compared with conventional iterative algorithms, D$^2$-ADMM has a faster convergence speed. Moreover, we proposed an effective monodirectional information exchange strategy to attain the cooperative design of all BSs with a small exchange overhead. Finally, numerical results demonstrated that the proposed D$^2$-ADMM achieve around $210\%$ improvement in capacity compared with the distributed noncooperative algorithm and almost $96\%$ compared with the centralized algorithm.

\ifCLASSOPTIONcaptionsoff
\newpage
\fi

\bibliographystyle{IEEEtran}
\bibliography{IEEEref}

\begin{thebibliography}{10}
\providecommand{\url}[1]{#1}
\csname url@samestyle\endcsname
\providecommand{\newblock}{\relax}
\providecommand{\bibinfo}[2]{#2}
\providecommand{\BIBentrySTDinterwordspacing}{\spaceskip=0pt\relax}
\providecommand{\BIBentryALTinterwordstretchfactor}{4}
\providecommand{\BIBentryALTinterwordspacing}{\spaceskip=\fontdimen2\font plus
\BIBentryALTinterwordstretchfactor\fontdimen3\font minus
  \fontdimen4\font\relax}
\providecommand{\BIBforeignlanguage}[2]{{%
\expandafter\ifx\csname l@#1\endcsname\relax
\typeout{** WARNING: IEEEtran.bst: No hyphenation pattern has been}%
\typeout{** loaded for the language `#1'. Using the pattern for}%
\typeout{** the default language instead.}%
\else
\language=\csname l@#1\endcsname
\fi
#2}}
\providecommand{\BIBdecl}{\relax}
\BIBdecl

\bibitem{B5G}
K.~Samdanis and T.~Taleb, ``The road beyond 5{G}: A vision and insight of the
  key technologies,'' \emph{IEEE Netw.}, vol.~34, no.~2, pp. 135--141, Feb.
  2020.

\bibitem{B5G2}
P.~Popovski, K.~F. Trillingsgaard, O.~Simeone, and G.~Durisi, ``5{G} wireless
  network slicing for e{MBB}, {URLLC}, and m{MTC}: A communication-theoretic
  view,'' \emph{IEEE Access}, vol.~6, pp. 55\,765--55\,779, Sep. 2018.

\bibitem{6G}
P.~Yang, Y.~Xiao, M.~Xiao, and S.~Li, ``6{G} wireless communications: Vision
  and potential techniques,'' \emph{IEEE Netw.}, vol.~33, no.~4, pp. 70--75,
  Aug. 2019.

\bibitem{hongbinli2}
J.~Mitola, J.~Guerci, J.~Reed, Y.-D. Yao, Y.~Chen, T.~C. Clancy, J.~Dwyer,
  H.~Li, H.~Man, R.~McGwier, and Y.~Guo, ``Accelerating {5G} {Q}o{E} via
  public-private spectrum sharing,'' \emph{IEEE Commun. Mag.}, vol.~52, no.~5,
  pp. 77--85, 2014.

\bibitem{smallcell}
S.~Hur, T.~Kim, D.~J. Love, J.~V. Krogmeier, T.~A. Thomas, and A.~Ghosh,
  ``Millimeter wave beamforming for wireless backhaul and access in small cell
  networks,'' \emph{IEEE Trans. Commun.}, vol.~61, no.~10, pp. 4391--4403, Oct.
  2013.

\bibitem{cellularMIMO}
C.-X. Wang, F.~Haider, X.~Gao, X.-H. You, Y.~Yang, D.~Yuan, H.~M. Aggoune,
  H.~Haas, S.~Fletcher, and E.~Hepsaydir, ``Cellular architecture and key
  technologies for 5{G} wireless communication networks,'' \emph{IEEE Commun.
  Mag.}, vol.~52, no.~2, pp. 122--130, Feb. 2014.

\bibitem{cellfree1}
H.~Q. Ngo, A.~Ashikhmin, H.~Yang, E.~G. Larsson, and T.~L. Marzetta,
  ``Cell-free massive {MIMO} versus small cells,'' \emph{IEEE Trans. Wireless
  Commun.}, vol.~16, no.~3, pp. 1834--1850, Mar. 2017.

\bibitem{comsmacel1}
W.~Liu, S.~Han, and C.~Yang, ``Energy efficiency comparison of massive {MIMO}
  and small cell network,'' in \emph{Proc. IEEE GlobalSIP}, Atlanta, GA, USA,
  Dec. 2014, pp. 617--621.

\bibitem{comsmacel2}
E.~Björnson, L.~Sanguinetti, and M.~Kountouris, ``Deploying dense networks for
  maximal energy efficiency: Small cells meet massive {MIMO},'' \emph{IEEE J.
  Sel. Areas Commun.}, vol.~34, no.~4, pp. 832--847, Apr. 2016.

\bibitem{comsmacel4}
H.~S. Dhillon, M.~Kountouris, and J.~G. Andrews, ``Downlink {MIMO} hetnets:
  Modeling, ordering results and performance analysis,'' \emph{IEEE Trans.
  Wireless Commun.}, vol.~12, no.~10, pp. 5208--5222, Oct. 2013.

\bibitem{comsmacel3}
S.~Zhou, M.~Zhao, X.~Xu, J.~Wang, and Y.~Yao, ``Distributed wireless
  communication system: a new architecture for future public wireless access,''
  \emph{IEEE Commun. Mag.}, vol.~41, no.~3, pp. 108--113, Mar. 2003.

\bibitem{cellfree2}
E.~Nayebi, A.~Ashikhmin, T.~L. Marzetta, H.~Yang, and B.~D. Rao, ``Precoding
  and power optimization in cell-free massive {MIMO} systems,'' \emph{IEEE
  Trans. Wireless Commun.}, vol.~16, no.~7, pp. 4445--4459, Jul. 2017.

\bibitem{cellfree3}
E.~Björnson and L.~Sanguinetti, ``Making cell-free massive {MIMO} competitive
  with {MMSE} processing and centralized implementation,'' \emph{IEEE Trans.
  Wireless Commun.}, vol.~19, no.~1, pp. 77--90, Jan. 2020.

\bibitem{cellfree4}
H.~Q. Ngo, A.~Ashikhmin, H.~Yang, E.~G. Larsson, and T.~L. Marzetta,
  ``Cell-free massive {MIMO}: Uniformly great service for everyone,'' in
  \emph{Proc. IEEE SPAWC}, Stockholm, Sweden, Jun. 2015, pp. 201--205.

\bibitem{highfreband1}
M.~Xiao, S.~Mumtaz, Y.~Huang, L.~Dai, Y.~Li, M.~Matthaiou, G.~K. Karagiannidis,
  E.~Björnson, K.~Yang, C.-L. I, and A.~Ghosh, ``Millimeter wave
  communications for future mobile networks,'' \emph{IEEE J. Sel. Areas
  Commun.}, vol.~35, no.~9, pp. 1909--1935, Sep. 2017.

\bibitem{highfreband2}
Z.~Chen, X.~Ma, B.~Zhang, Y.~Zhang, Z.~Niu, N.~Kuang, W.~Chen, L.~Li, and
  S.~Li, ``A survey on terahertz communications,'' \emph{China Commun.},
  vol.~16, no.~2, pp. 1--35, Feb. 2019.

\bibitem{1WQQtoward2020}
Q.~{Wu} and R.~{Zhang}, ``Towards smart and reconfigurable environment:
  Intelligent reflecting surface aided wireless network,'' \emph{IEEE Commun.
  Mag.}, vol.~58, no.~1, pp. 106--112, Jan. 2020.

\bibitem{huangTWC}
C.~Huang, A.~Zappone, G.~C. Alexandropoulos, M.~Debbah, and C.~Yuen,
  ``Reconfigurable intelligent surfaces for energy efficiency in wireless
  communication,'' \emph{IEEE Trans. Wireless Commun.}, vol.~18, no.~8, pp.
  4157--4170, Aug. 2019.

\bibitem{xuwangyang}
W.~Xu, L.~Gan, and C.~Huang, ``A robust deep learning-based beamforming design
  for {RIS}-assisted multiuser {MISO} communications with practical
  constraints,'' \emph{IEEE Trans. Cogn. Commun. Netw.}, vol.~8, no.~2, pp.
  694--706, Jun. 2022.

\bibitem{an3}
J.~An, C.~Xu, L.~Gan, and L.~Hanzo, ``Low-complexity channel estimation and
  passive beamforming for {RIS}-assisted {MIMO} systems relying on discrete
  phase shifts,'' \emph{IEEE Trans. Commun.}, vol.~70, no.~2, pp. 1245--1260,
  Feb. 2022.

\bibitem{xuwangyang1}
W.~Xu, J.~An, C.~Huang, L.~Gan, and C.~Yuen, ``Deep reinforcement learning
  based on location-aware imitation environment for {RIS}-aided mmwave {MIMO}
  systems,'' \emph{IEEE Wireless Commun. Lett.}, vol.~11, no.~7, pp.
  1493--1497, Jul. 2022.

\bibitem{cellfreenoncopZF}
S.~Buzzi, C.~D’Andrea, A.~Zappone, and C.~D’Elia, ``User-centric 5{G}
  cellular networks: Resource allocation and comparison with the cell-free
  massive {MIMO} approach,'' \emph{IEEE Trans. Wireless Commun.}, vol.~19,
  no.~2, pp. 1250--1264, Feb. 2020.

\bibitem{cellfreenoncopMMSE}
M.~Alonzo, S.~Buzzi, A.~Zappone, and C.~D’Elia, ``Energy-efficient power
  control in cell-free and user-centric massive {MIMO} at millimeter wave,''
  \emph{IEEE Trans. Green Commun. Netw.}, vol.~3, no.~3, pp. 651--663, Sep.
  2019.

\bibitem{cellfreecopZF1}
E.~Nayebi, A.~Ashikhmin, T.~L. Marzetta, H.~Yang, and B.~D. Rao, ``Precoding
  and power optimization in cell-free massive {MIMO} systems,'' \emph{IEEE
  Trans. Wireless Commun.}, vol.~16, no.~7, pp. 4445--4459, Jul. 2017.

\bibitem{cellfreecopZF2}
P.~Liu, K.~Luo, D.~Chen, and T.~Jiang, ``Spectral efficiency analysis of
  cell-free massive {MIMO} systems with zero-forcing detector,'' \emph{IEEE
  Trans. Wireless Commun.}, vol.~19, no.~2, pp. 795--807, Feb. 2020.

\bibitem{hongbinli1}
J.~Wang, B.~Wang, J.~Fang, and H.~Li, ``Millimeter wave cell-free massive
  {MIMO} systems: Joint beamforming and ap-user association,'' \emph{IEEE
  Wireless Commun. Lett.}, vol.~11, no.~2, pp. 298--302, 2022.

\bibitem{cellfreecd1}
A.~Tolli, H.~Ghauch, J.~Kaleva, P.~Komulainen, M.~Bengtsson, M.~Skoglund,
  M.~Honig, E.~Lahetkangas, E.~Tiirola, and K.~Pajukoski, ``Distributed
  coordinated transmission with forward-backward training for 5{G} radio
  access,'' \emph{IEEE Commun. Mag.}, vol.~57, no.~1, pp. 58--64, Jan. 2019.

\bibitem{cellfreecd2}
J.~Kaleva, A.~Tölli, M.~Juntti, R.~A. Berry, and M.~L. Honig, ``Decentralized
  joint precoding with pilot-aided beamformer estimation,'' \emph{IEEE Trans.
  Signal Process.}, vol.~66, no.~9, pp. 2330--2341, May 2018.

\bibitem{wujointcontinuous}
Q.~Wu and R.~Zhang, ``Intelligent reflecting surface enhanced wireless network
  via joint active and passive beamforming,'' \emph{IEEE Trans. Wireless
  Commun.}, vol.~18, no.~11, pp. 5394--5409, Nov. 2019.

\bibitem{an1}
J.~An and L.~Gan, ``The low-complexity design and optimal training overhead for
  {IRS}-assisted {MISO} systems,'' \emph{IEEE Wireless Commun. Lett.}, vol.~10,
  no.~8, pp. 1820--1824, Aug. 2021.

\bibitem{huang2020reconfigurable}
C.~Huang, R.~Mo, and C.~Yuen, ``Reconfigurable intelligent surface assisted
  multiuser {MISO} systems exploiting deep reinforcement learning,'' \emph{IEEE
  J. Sel. Areas Commun.}, vol.~38, no.~8, pp. 1839--1850, Oct. 2020.

\bibitem{huang2020holographic}
C.~Huang, S.~Hu, G.~C. Alexandropoulos, A.~Zappone, C.~Yuen, R.~Zhang,
  M.~Di~Renzo, and M.~Debbah, ``Holographic {{MIMO}} surfaces for 6{G} wireless
  networks: Opportunities, challenges, and trends,'' \emph{IEEE Wireless
  Commun.}, vol.~27, no.~5, pp. 118--125, Oct. 2020.

\bibitem{RIScellfree1}
E.~Shi, J.~Zhang, S.~Chen, J.~Zheng, Y.~Zhang, D.~W. Kwan~Ng, and B.~Ai,
  ``Wireless energy transfer in {RIS}-aided cell-free massive {MIMO} systems:
  Opportunities and challenges,'' \emph{IEEE Commun. Mag.}, vol.~60, no.~3, pp.
  26--32, Mar. 2022.

\bibitem{RIScellfree2}
B.~Al-Nahhas, M.~Obeed, A.~Chaaban, and M.~J. Hossain, ``{RIS}-aided cell-free
  massive {MIMO}: Performance analysis and competitiveness,'' in \emph{Proc.
  IEEE ICC Workshops}, Montreal, QC, Canada, Jun. 2021, pp. 1--6.

\bibitem{RIScellfree3}
Q.~N. Le, V.-D. Nguyen, O.~A. Dobre, and R.~Zhao, ``Energy efficiency
  maximization in {RIS}-aided cell-free network with limited backhaul,''
  \emph{IEEE Commun. Lett.}, vol.~25, no.~6, pp. 1974--1978, Jun. 2021.

\bibitem{RIScellfree4}
Z.~Zhang and L.~Dai, ``A joint precoding framework for wideband reconfigurable
  intelligent surface-aided cell-free network,'' \emph{IEEE Trans. Signal
  Process.}, vol.~69, pp. 4085--4101, Jun. 2021.

\bibitem{RIScellfree5}
S.~Huang, Y.~Ye, M.~Xiao, H.~V. Poor, and M.~Skoglund, ``Decentralized
  beamforming design for intelligent reflecting surface-enhanced cell-free
  networks,'' \emph{IEEE Wireless Commun. Lett.}, vol.~10, no.~3, pp. 673--677,
  Mar. 2021.

\bibitem{antgcn}
J.~An, C.~Xu, L.~Wang, Y.~Liu, L.~Gan, and L.~Hanzo, ``Joint training of the
  superimposed direct and reflected links in reconfigurable intelligent surface
  assisted multiuser communications,'' \emph{IEEE Trans. Green Commun. Netw.},
  vol.~6, no.~2, pp. 739--754, June 2022.

\bibitem{anwc}
J.~An, C.~Xu, Q.~Wu, D.~W.~K. Ng, M.~D. Renzo, C.~Yuen, and L.~Hanzo,
  ``Codebook-based solutions for reconfigurable intelligent surfaces and their
  open challenges,'' \emph{IEEE Wireless Commun.}, pp. 1--8, 2022, Early
  Access.

\bibitem{xu3}
W.~Xu, J.~An, Y.~Xu, C.~Huang, L.~Gan, and C.~Yuen, ``Time-varying channel
  prediction for {RIS}-assisted {MU-MISO} networks via deep learning,''
  \emph{IEEE Trans. Cogn. Commun. Netw.}, vol.~8, no.~4, pp. 1802--1815, 2022.

\bibitem{deepphy}
H.~Huang, S.~Guo, G.~Gui, Z.~Yang, J.~Zhang, H.~Sari, and F.~Adachi, ``Deep
  learning for physical-layer 5{G} wireless techniques: Opportunities,
  challenges and solutions,'' \emph{IEEE Wireless Commun.}, vol.~27, no.~1, pp.
  214--222, Feb. 2020.

\bibitem{Alunroll1}
V.~Monga, Y.~Li, and Y.~C. Eldar, ``Algorithm unrolling: Interpretable,
  efficient deep learning for signal and image processing,'' \emph{IEEE Signal
  Process. Mag.}, vol.~38, no.~2, pp. 18--44, Mar. 2021.

\bibitem{Alunroll2}
B.~Xin, Y.~Wang, W.~Gao, D.~Wipf, and B.~Wang, ``Maximal sparsity with deep
  networks?'' \emph{Adv. in Neural Inf. Process. Syst.}, vol.~29, 2016.

\bibitem{Alunroll3}
J.~Liu and X.~Chen, ``Alista: Analytic weights are as good as learned weights
  in lista,'' in \emph{Proc. ICLR}, 2019.

\bibitem{Alunroll4}
X.~Chen, J.~Liu, Z.~Wang, and W.~Yin, ``Theoretical linear convergence of
  unfolded ista and its practical weights and thresholds,'' \emph{Advances in
  Neural Information Processing Systems}, vol.~31, 2018.

\bibitem{Alunroll5}
Y.~Yang, J.~Sun, H.~Li, and Z.~Xu, ``{ADMM-CSN}et: A deep learning approach for
  image compressive sensing,'' \emph{IEEE Trans. Pattern Anal. Mach. Intell.},
  vol.~42, no.~3, pp. 521--538, Mar. 2020.

\bibitem{Alunroll6}
S.~A.~H. Hosseini, B.~Yaman, S.~Moeller, M.~Hong, and M.~Akçakaya, ``Dense
  recurrent neural networks for accelerated mri: History-cognizant unrolling of
  optimization algorithms,'' \emph{IEEE J. Sel. Topics Signal Process.},
  vol.~14, no.~6, pp. 1280--1291, Oct. 2020.

\bibitem{Alunroll7}
Y.~Li, M.~Tofighi, J.~Geng, V.~Monga, and Y.~C. Eldar, ``Efficient and
  interpretable deep blind image deblurring via algorithm unrolling,''
  \emph{IEEE Trans. Comput. Imag.}, vol.~6, pp. 666--681, Jan. 2020.

\bibitem{Alunroll8}
X.~Zhang, Y.~Lu, J.~Liu, and B.~Dong, ``Dynamically unfolding recurrent
  restorer: A moving endpoint control method for image restoration,''
  \emph{arXiv preprint arXiv:1805.07709}, 2018.

\bibitem{Alunroll9}
J.~R. Hershey, J.~L. Roux, and F.~Weninger, ``Deep unfolding: Model-based
  inspiration of novel deep architectures,'' \emph{arXiv preprint
  arXiv:1409.2574}, 2014.

\bibitem{Alunroll10}
S.~Lohit, D.~Liu, H.~Mansour, and P.~T. Boufounos, ``Unrolled projected
  gradient descent for multi-spectral image fusion,'' in \emph{Proc. IEEE
  ICASSP}, Brighton, UK, May 2019, pp. 7725--7729.

\bibitem{wangSVmodel}
P.~Wang, J.~Fang, H.~Duan, and H.~Li, ``Compressed channel estimation for
  intelligent reflecting surface-assisted millimeter wave systems,'' \emph{IEEE
  Signal Process. Lett.}, vol.~27, pp. 905--909, May 2020.

\bibitem{LT2}
K.~Shen and W.~Yu, ``Fractional programming for communication systems—part
  {II}: Uplink scheduling via matching,'' \emph{IEEE Trans. Signal Process.},
  vol.~66, no.~10, pp. 2631--2644, May 2018.

\bibitem{LT1}
K.~Shen and W.~Yu, ``Fractional programming for communication systems—part
  {I}: Power control and beamforming,'' \emph{IEEE Trans. Signal Process.},
  vol.~66, no.~10, pp. 2616--2630, May 2018.

\bibitem{ADMM1}
Y.~Ye, H.~Chen, Z.~Ma, and M.~Xiao, ``Decentralized consensus optimization
  based on parallel random walk,'' \emph{IEEE Commun. Lett.}, vol.~24, no.~2,
  pp. 391--395, Feb. 2020.

\bibitem{ADMM2}
X.~Yu, J.-C. Shen, J.~Zhang, and K.~B. Letaief, ``Alternating minimization
  algorithms for hybrid precoding in millimeter wave {MIMO} systems,''
  \emph{IEEE J. Sel. Topics Signal Process.}, vol.~10, no.~3, pp. 485--500,
  Apr. 2016.

\bibitem{LocalZF}
G.~Interdonato, M.~Karlsson, E.~Björnson, and E.~G. Larsson, ``Local partial
  zero-forcing precoding for cell-free massive {MIMO},'' \emph{IEEE Trans.
  Wireless Commun.}, vol.~19, no.~7, pp. 4758--4774, Jul. 2020.

\end{thebibliography}
\end{document}